\documentclass[preprint,amsmath,amssymb,nofootinbib]{revtex4}
\usepackage[dvips]{graphics}
\usepackage{graphics}

\usepackage{multirow}
\usepackage{amsmath}
\usepackage{array}
\usepackage{booktabs}
\makeatletter
\renewcommand{\fnum@figure}{Fig.~ \thefigure}
\makeatother
\usepackage{pstcol}
\usepackage{amsfonts}
\usepackage{bm}
\usepackage{amsmath}
\usepackage{comment}
\usepackage{amssymb}
\usepackage{color}
\usepackage[all]{xy}
\usepackage{comment}
\usepackage{epstopdf,epsfig}
\def\be{\begin{equation}}
\def\ee{\end{equation}}
\def\bea{\begin{eqnarray}}
\def\eea{\end{eqnarray}}
\usepackage{float}
\usepackage{xcolor}
\usepackage{xcolor}
\definecolor{fucsia}{HTML}{FF00FF}
\usepackage{graphicx}
\usepackage{makecell}
\usepackage{multirow}
\usepackage{subcaption}
\usepackage{setspace}
\usepackage{microtype}
\usepackage[
    colorlinks=true,
    linkcolor=blue,     % Color de los enlaces en TOC, ecuaciones, figuras, etc.
    citecolor=red,      % Color de referencias bibliográficas
]{hyperref}

\usepackage{ragged2e}

\makeatletter
\def\@makecaption#1#2{%
  \vskip 8pt % <-- espacio entre figura y caption (ajusta aquí)
  \begingroup
    \small
    \justifying
    \textbf{#1.} #2\par
  \endgroup
}
\makeatother
\usepackage{scalerel}
\usepackage{tikz}
\usetikzlibrary{svg.path}

\definecolor{orcidlogocol}{HTML}{A6CE39}
\tikzset{
  orcidlogo/.pic={
    \fill[orcidlogocol] svg{M256,128c0,70.7-57.3,128-128,128C57.3,256,0,198.7,0,128C0,57.3,57.3,0,128,0C198.7,0,256,57.3,256,128z};
    \fill[white] svg{M86.3,186.2H70.9V79.1h15.4v48.4V186.2z}
                 svg{M108.9,79.1h41.6c39.6,0,57,28.3,57,53.6c0,27.5-21.5,53.6-56.8,53.6h-41.8V79.1z M124.3,172.4h24.5c34.9,0,42.9-26.5,42.9-39.7c0-21.5-13.7-39.7-43.7-39.7h-23.7V172.4z}
                 svg{M88.7,56.8c0,5.5-4.5,10.1-10.1,10.1c-5.6,0-10.1-4.6-10.1-10.1c0-5.6,4.5-10.1,10.1-10.1C84.2,46.7,88.7,51.3,88.7,56.8z};
  }
}

\newcommand\orcidicon[1]{\href{https://orcid.org/#1}{\mbox{\scalerel*{
\begin{tikzpicture}[yscale=-1,transform shape]
\pic{orcidlogo};
\end{tikzpicture}
}{|}}}}
\begin{document}
\baselineskip=20pt

\title{Geometric Universality and Thermodynamic Microstructure of Real Fluids in a Unified Entropic Framework}

\author{Carlos E. Romero-Figueroa\textsuperscript{\orcidicon{0000-0001-5548-7766}1}, Jose Miguel Ladino\textsuperscript{\orcidicon{0000-0001-9812-4949}1}, Sasha A. Zaldivar\textsuperscript{\orcidicon{0009-0009-9350-055X}1}, Hernando Quevedo\textsuperscript{\orcidicon{0000-0003-4433-5550}1,2,3}}
\email{\raggedright
carlosed.romero@correo.nucleares.unam.mx;
miguel.ladino@correo.nucleares.unam.mx; sasha@ciencias.unam.mx;
quevedo@nucleares.unam.mx
}
\affiliation{\hspace{1cm}\\ \mbox{$^1$Instituto~de~Ciencias~Nucleares,~Universidad~Nacional~Aut\'onoma~de~M\'exico,}\\ \mbox{AP~70543,~Mexico~City,~Mexico}}
\affiliation{\mbox{$^2$Dipartimento~di~Fisica~and~Icra,~Universit\`a~di~Roma~“La~Sapienza”,~Roma,~Italy}}
\affiliation{\mbox{$^3$Al-Farabi~Kazakh~National~University,~Al-Farabi~av.~71,~050040~Almaty,~Kazakhstan}\\}

\date{\today}
\begin{abstract}
We introduce a unified entropic framework for real fluids that encompasses the van der Waals, Berthelot, Redlich–Kwong, and Peng–Robinson equations of state within a common thermodynamic description. The corresponding microscopic interactions are then explored using Geometrothermodynamics (GTD) through the scalar curvature $\mathcal{R}$ of the equilibrium manifold. We show that curvature singularities accurately reproduce macroscopic critical phenomena, while vanishing curvature ($\mathcal{R}=0$) identifies specific thermodynamic states where attractive and repulsive intermolecular forces effectively balance. Furthermore, we introduce a set of dimensionless critical-amplitude ratios $Q^i_{\;j}$, which reveal universal geometric features of the critical regime. Although individual critical amplitudes exhibit a logarithmic dependence on the system size, these invariant ratios organize different molecular species according to the strength of criticality and encode universal scaling features, suggesting their potential as robust classification parameters. Finally, employing Bayesian inference and Markov Chain Monte Carlo (MCMC) methods, we statistically reconstruct the zero-curvature curves. The posterior distributions support the consistency of the geometric scaling behavior, demonstrating that the GTD manifold encodes non-trivial information about the underlying thermodynamical models.

\vspace{0.1in}

\end{abstract}

\maketitle
\section{Introduction}
The study of real fluids is a fundamental component of thermodynamics, frequently serving to align theoretical idealizations with the empirical behavior and intermolecular interactions of actual physical systems. Unlike ideal gases, which assume non-interacting point particles, real fluids are modeled by a interplay of microscopic forces, namely, long-range intermolecular attractions and short-range repulsions due to finite molecular volume. These underlying interactions are responsible for the rich phenomenological landscape observed in real systems, including phase transitions and the emergence of critical points. Understanding how these microscopic details manifest macroscopically is crucial not only for fundamental physics but also for diverse engineering and technological applications, driving the continuous search for robust theoretical frameworks capable of capturing the intricate dynamics of fluids across different thermodynamic regimes. Historically, the quest to accurately model these phenomena has been driven by the development of increasingly sophisticated Equations of State (EoS). In 1873, J. D. van der Waals modified the ideal gas law to account for finite molecular size (excluded volume) and attractive intermolecular forces \cite{van_der_Waals_1873}. While this model successfully predicted liquid-vapor phase transitions and the existence of critical phenomena, its severe quantitative limitations, particularly regarding liquid densities and behavior in the critical region, prompted further empirical developments \cite{Valderrama_2003}. In 1899, D. Berthelot modified the van der Waals (vdW) model by introducing a temperature-dependent attractive term, which improved the representation of compressibility at low pressures, though it remained inaccurate near the critical state \cite{Berthelot_1899, Poling_2001}. In 1949, O. Redlich and J.N.S. Kwong introduced a more robust temperature dependence in the attractive term, significantly improving the representation of gas-phase properties at high pressures over a wide temperature range \cite{Redlich_Kwong_1949}. To address persistent inaccuracies in liquid density predictions and critical compressibility factors inherent in earlier models, D. Y. Peng and D. B. Robinson developed the Peng–Robinson equation in 1976 \cite{Peng_Robinson_1976}. This model quickly became a chemical engineering standard for accurately modeling vapor-liquid equilibrium, particularly for hydrocarbon mixtures \cite{Sandler_2017}. An extensive review of research prior to 2000, \cite{Valderrama_2003} highlights how modifying the temperature function $\alpha(T)$ of Cubic Equations of State (CEoS) improves polar fluid vapor pressure predictions. Indeed, a vast landscape of generalized formulations significantly enhances the thermodynamic accuracy of CEoS in modeling real fluid behavior \cite{Valderrama_2003, Trummler, NASRIFAR200173, DUAN20042997, KISELEV19987, GURIA201279, ASHOUR200041, ESMAEILZADEH200683, KIM20121351, Kumar82, Pedersen, Ali2024, MOHSENNIA200327, Wilczek, WeiWei, Sokovnin2022, Economou, SANCHEZ2023121167}. Despite differing functional forms, all these models describe macroscopic manifestation of microscopic interactions. \\

Geometrothermodynamics (GTD) provides a Legendre-invariant framework that endows the thermodynamic equilibrium space with a Riemannian manifold structure \cite{quevedo2007geometrothermodynamics,quevedo2023unified,bravetti2017zeroth}. Within this geometric description, phase structures and microscopic interactions are encoded in the curvature of the manifold. Curvature singularities indicate thermodynamic critical behavior, while the scalar curvature  quantifies the interaction regimes: positive, negative, and vanishing curvatures correspond to repulsive, attractive, and ideal-gas behaviors, respectively. Consequently, mapping the zero-curvature curves allows for the geometric analysis of the Boyle temperature, the specific temperature at which a real fluid behaves like an ideal gas. This work introduces a unified entropic framework to incorporate real fluid EoS into an analytically GTD representation. We investigate the resulting phase structures, critical phenomena, and zero-curvature curves to evaluate whether universal geometric features persist across different models. By validating the GTD through the well-established phenomenology of laboratory fluids, we construct a reliable baseline for broader physical interpretation. This is crucial for extending GTD insights to the diverse range of complex systems where the formalism has already been applied, including ideal and real gases \cite{quevedo2022geometrothermodynamics,quevedo2011phase,quevedo2023geometrothermodynamic}, ideal quantum gases \cite{zaldivar2023ideal}, magnetic materials \cite{quevedo2024geometrothermodynamic}, and the Ising model \cite{bravetti2014representation}. Its applicability further extends to interdisciplinary domains such as chemical reactions \cite{quevedo2014geometric} and econophysics \cite{2023IJGMM..2050057Q}, as well as cosmological models \cite{romero2026quasi,e25060944,PhysRevD.86.063508,2019EPJC...79..577B,2014GrCo...20..208Q}. Moreover, GTD has been extensively employed in the study of black hole thermodynamics across a wide variety of gravitational theories \cite{quevedo2008geometrothermodynamics,Ladino:2024ned,ladino2025phase,Larranga2011,Gogoi,larranaga2012geometric,ref54,ref39,Tharanath_2015,ZHANG2018170,Channuie:2018mkt,2025EPJC...85..785G,2018GReGr..50...20K,Sanchez:2016ger,Quevedo:2019wbz,2017GReGr..49..148H,2016ChPhB..25l0401G,2025IJGMM..2250049J,2025PDU....5002146S,Taj:2012sir,Mo:2013qhv,Quevedo_2009,Luciano_2023,Luciano_20232,2023EPJC...83..710R,2024Entrp..26..457B,Janke_2010,Quevedo_2016,Hendi_2015,Han_2012,Bravetti_2013,Quevedo:2011fk,Rizwan:2018ozh,Quevedo:2016swn,ahmed2026shadow,romero2024extended1,ladino2026probing,quevedo2019quasi}.\\

The paper is organized as follows: Section~\ref{sec:II} establishes the generalized entropic representation and the EoS of real fluids under consideration, and details the virial coefficients and Boyle temperature formulations. In Section~\ref{GTD}, we investigate the geometrothermodynamics and microstructure with a special emphasis on the van der Waals case. In Section~\ref{universa}, we extend this analysis to the Berthelot and Redlich–Kwong models, extracting their universal critical exponents and characterizing their geometric universality near criticality. In Section~\ref{Zero curv}, we explore the zero-curvature curves and perform a robust Bayesian Markov Chain Monte Carlo (MCMC) analysis to infer the asymptotic power-law scaling of the zero-curvature temperature. Finally, Section~\ref{sec:VI} contains our concluding remarks.

\section{General Entropic Representation of Real Fluids}
\label{sec:II}
The thermodynamics of a real fluid derives from the canonical partition function $Z(N,V,T)$ and configuration integral $Q_N(V,T)$
\begin{equation}
Z(N,V,T) = \frac{1}{N! \Lambda^{3N}} Q_N(V,T), \qquad Q_N(V,T) = \int e^{-\beta U(\vec{r})} d\vec{r},
\end{equation}
where $T$ denotes the temperature, $V$ the volume, $N$ the number of particles, $\Lambda$ is the thermal de Broglie wavelength, $\vec{r} \equiv \{\vec{r}_1, \dots, \vec{r}_N\}$ are the particles' spatial coordinates, and $\beta = 1/(k_B T)$. For real fluids, a mean-field approximation decouples the intermolecular potential $U(\vec{r})$ into long-range attraction $U_{attr}$ and short-range repulsion, approximating the integral as
\begin{equation}
Q_N(V,T) \approx (V - N b)^N e^{-\beta U_{attr}},
\end{equation}
where $(V - Nb)^N$ accounts for the excluded free volume from repulsive forces, with $b$ as the effective covolume. By applying Stirling's approximation to $F = -k_B T \ln Z$, the repulsive interactions yield the standard free-volume contribution $\ln(V/N - b)$. To determine the attractive contribution $U_{attr}$ from an empirical EoS via reverse thermodynamics \cite{Sandler_2017}, we extend this framework for real fluids by modeling $U_{attr}$ using two functions: $A(V,T,N)$ and $\Theta(T)$. Under this approach, the generalized Helmholtz free energy becomes
\begin{align}
F &= - S_0 T - \frac{1}{2} k_B N T \Big[ -3 + 3 \ln\!\left(\frac{3 k_B T}{2}\right) + 2 \ln\!\left(\frac{V}{N} - b\right) \Big] - A - \frac{T}{1+\Theta} \left(\frac{\partial A}{\partial T}\right)_{V,N}. \label{Free energy generalized}
\end{align}
where $S_0$ represents the reference entropy choice, which contains terms associated with $\Lambda$. The thermodynamic quantities are then obtained from
\begin{equation}
dF = -S\,dT - P\,dV + \mu\,dN.
\label{first law}
\end{equation}
Thus, the entropy $S$, the pressure $P$, and chemical potential $\mu$ follows
\begin{align}
S &= -\left(\frac{\partial F}{\partial T}\right)_{V,N}, \qquad P = -\left(\frac{\partial F}{\partial V}\right)_{T,N}, \qquad \mu = \left(\frac{\partial F}{\partial N}\right)_{T,V}.
\end{align}
The entropy of the system is
\begin{align}
S &= S_0 + k_B N \left[ \frac{3}{2} \ln\!\left(\frac{3 k_B T}{2}\right) + \ln\!\left(\frac{V}{N} - b\right) \right] \notag \\
&\quad + \frac{\left(2 + \Theta\big(3 + \Theta\big) - T\frac{d\Theta}{dT}\right)}{(1+\Theta)^2} \left(\frac{\partial A}{\partial T}\right)_{V,N} + \frac{T}{1+\Theta} \left(\frac{\partial^2 A}{\partial T^2}\right)_{V,N}.
\label{firstentropy}
\end{align}
For the pressure, we obtain the generalized EoS as
\begin{equation}
P = \frac{k_B NT}{V-Nb} + \left(\frac{\partial A}{\partial V}\right)_{T,N} + \frac{T}{1+\Theta} \left(\frac{\partial^2 A}{\partial T\,\partial V}\right)_{N},
\label{eq;EoSGeneral}
\end{equation}
and the chemical potential is
\begin{align}
\mu &= \frac{1}{2} k_B T \left[ 5 + \frac{2 b N}{V - b N} - 3 \ln\!\left(\frac{3 k_B T}{2}\right) - 2 \ln\!\left(\frac{V}{N} - b\right) \right] - \left(\frac{\partial A}{\partial N}\right)_{V,T} - \frac{T}{1+\Theta} \left(\frac{\partial^2 A}{\partial N\,\partial T}\right)_{V}.
\end{align}
Therefore, the first law of thermodynamics together with the corresponding Maxwell relation are
\begin{align}
dU &= T dS - P dV + \mu dN, \qquad \text{and} \qquad \left( \frac{\partial S}{\partial V} \right)_{T,N} = \left( \frac{\partial P}{\partial T} \right)_{V,N}.
\label{firstlaw}
\end{align}
Where $U$ denotes the internal energy. In this framework, the internal energy can be written as
\begin{align}
U(V,T,N) &= \frac{3}{2} N k_B T - A(V,T,N), \qquad A(V,T,N) \equiv \int_{V}^{\infty} \left[ T \left( \frac{\partial P}{\partial T}  \right)_{V,N}-P \right] dV,
\label{GeneralEnergy}
\end{align}
where the function $A$ encodes the deviations from ideal behavior due to intermolecular interactions. To ensure absolute thermodynamic consistency, the functions $A$ and $\Theta$ must satisfy the following identity
\begin{equation}
\frac{(1+\Theta)^2 - T \frac{d\Theta}{dT}}{1+\Theta} \left( \frac{\partial A}{\partial T} \right)_{V,N} + T \left( \frac{\partial^2 A}{\partial T^2} \right)_{V,N} = 0.
\label{eq:identity}
\end{equation}
This condition guarantees the correct physical coupling between $A$ and $\Theta$, ensuring that the internal energy $U$ defined in Eq.~\eqref{GeneralEnergy} remains strictly consistent with the fundamental Helmholtz relation, $F = U - TS$. By satisfying this identity, the proposed method adheres entirely to the formal structure of the laws of thermodynamics, the Maxwell relations, and the necessary stability criteria of the state functions. Consequently, from Eqs.~\eqref{firstentropy}, \eqref{GeneralEnergy}, and \eqref{eq:identity}, we obtain the generalized entropy functional compatible with this approach
\begin{align}
S(U,V,N) &= S_0 + k_B N \left[ \frac{3}{2} \ln\!\left(\frac{U + A}{N}\right) + \ln\!\left(\frac{V}{N} - b\right) \right] + \frac{1}{1+\Theta} \left(\frac{\partial A}{\partial T}\right)_{V,N}.
\label{GeneralEntropySimple}
\end{align}

\subsection{Real Fluid Equations of State}
\begin{figure}[ht!]
\centering

\begin{minipage}{0.49\textwidth}
    \centering
    \includegraphics[width=\linewidth]{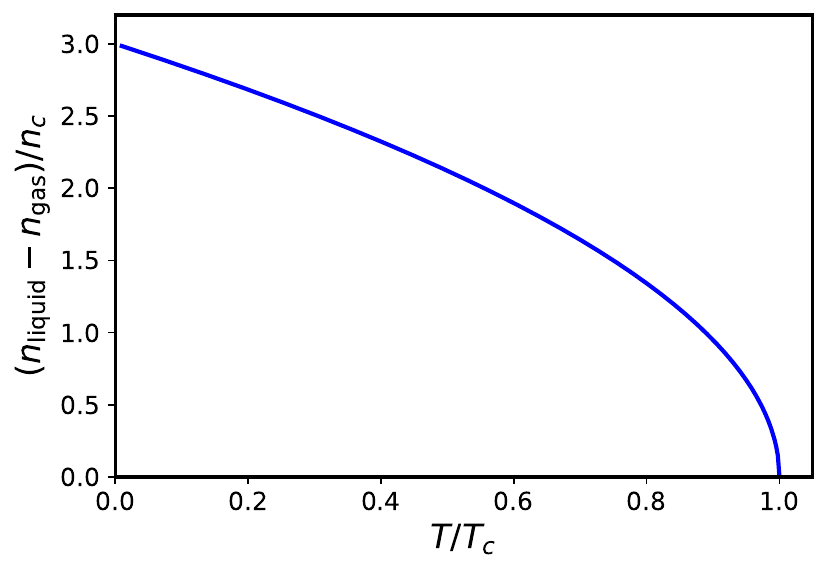}
\end{minipage}
\hfill
\begin{minipage}{0.498\textwidth}
    \centering
    \includegraphics[width=\linewidth]{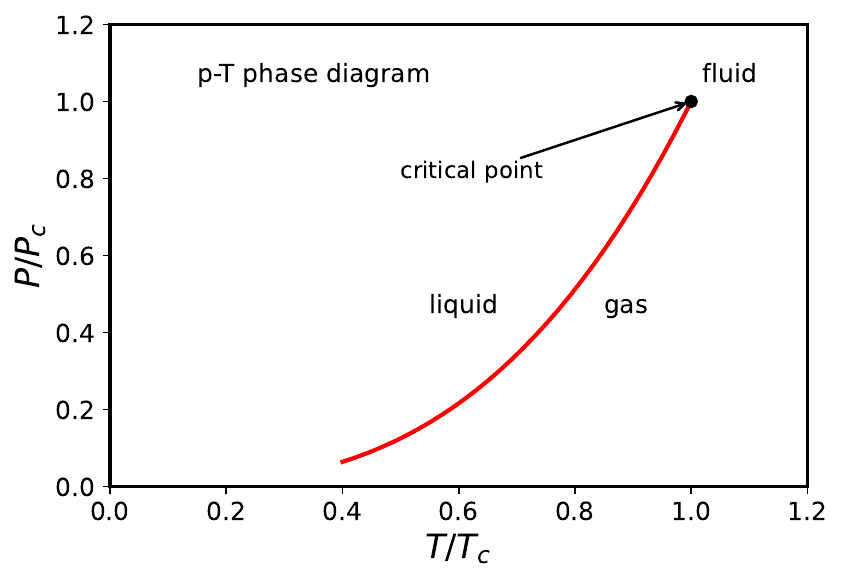}
\end{minipage}

\caption{
(a) Order parameter as a function of $T/T_c$ for the vdW fluid.
(b) Phase diagram in the reduced $(P/P_c,\,T/T_c)$ plane.
}
\label{fig:vdw_coexistence}
\end{figure}

Real fluids are widely modeled using CEoS by partitioning the pressure into repulsive and attractive terms: $P = P_{\text{rep}} + P_{\text{attr}}$. While the repulsive part corrects the ideal gas law for the excluded volume $b$, the attractive term is scaled by a temperature-dependent function $\alpha(T)$ \cite{Valderrama_2003}. Given that $P_{\text{attr}}$ is separable into $\alpha(T)$ and a purely function of $(V, N)$, we can characterize the thermal consistency of the EoS (when $\Theta(T) \neq 0$) through the dimensionless parameter relation
\begin{align}
\Theta(T)
=
-1
-
T\frac{d^2\alpha(T)}{dT^2}
\left(\frac{d\alpha(T)}{dT}\right)^{-1}.
\label{Bparameter}
\end{align}
Thus, different CEoS of real fluids are recovered through specific choices of $A$ and $\Theta$, such as
{\small
\begin{align}
&&\textbf{van der Waals: }&
\quad
A = \frac{a N^2}{V},
&&
\Theta = 0
\label{eq:vdW}
\\
&&\textbf{Berthelot: }&
\quad
A = \frac{2 a N^2}{T V},
&&
\Theta = 1
\label{eq:berthelot}
\\
&&\textbf{Redlich-Kwong: }&
\quad
A = \frac{3 a N \ln\left( \frac{V+N b}{V}\right)}{2 b \sqrt{T}},
&&
\Theta = \frac{1}{2}
\label{eq:rk}
\\
&&\textbf{Peng-Robinson: }&
\quad
A =
\frac{a(1+m)N \alpha_{PR}(T)
\ln\left[\frac{V + (1+\sqrt{2})N b}
{V + (1-\sqrt{2})N b}\right]}
{2\sqrt{2}b},
&&
\Theta =
-1
-
T\frac{d^2\alpha_{PR}}{dT^2}
\left(\frac{d\alpha_{PR}}{dT}\right)^{-1}.
\label{eq:pr}
\end{align}
}
Here, $\alpha_{PR}(T) = \left[1 + m ( 1 - \sqrt{T/T_c} )\right]^2$ and $m = 0.37464 + 1.54226\omega - 0.26992\omega^2$, where $T_c$ is the critical temperature and $\omega$ is the acentric factor. From now on, the cases where $A=A(V,N)$ and $\Theta=0$ will be referred to as van der Waals-like cases. In contrast, more general scenarios with $A=A(V,N,T)$ and $\Theta(T)\neq0$ will be referred to as modified van der Waals cases. Therefore, the generalized EoS Eq.~\eqref{eq;EoSGeneral} can be specified as
\begin{align}
\textbf{vdW-like: } &\quad
P(V,N,T)
=
\frac{Nk_BT}{V-Nb}
+
\left(\frac{\partial A}{\partial V}\right)_{T,N},
\\[8pt]
\textbf{modified vdW: } &\quad
P(V,N,T)
=
\frac{Nk_BT}{V-Nb}
+
\left(\frac{\partial A}{\partial V}\right)_{T,N}
+
\frac{T}{1+\Theta}
\left(\frac{\partial^2 A}{\partial T\,\partial V}\right)_{N}.
\end{align}
As a consistency check, the ideal gas is recovered for $A=a=b=0$. Moreover, substituting $A$ and $\Theta$ from Eqs.~\eqref{eq:vdW}--\eqref{eq:pr} into the generalized EoS yields the corresponding real fluid EoS
{\small
\begin{align}
&&\textbf{van der Waals: }&
\quad
P(V,N,T)=\frac{ Nk_BT}{V-b N}-\frac{a N^2}{V^2},
\label{EoS vdW}
\\
&&\textbf{Berthelot: }&
\quad
P(V,N,T)=\frac{ Nk_BT}{V-b N}-\frac{a N^2}{T V^2},
\\
&&\textbf{Redlich-Kwong: }&
\quad
P(V,N,T)=\frac{ Nk_BT}{V-b N}-\frac{a N^2}{\sqrt{T} V (V+b N)},
\\
&&\textbf{Peng-Robinson: }&
\quad
P(V,N,T)=\frac{ Nk_BT}{V-b N}
-\frac{a N^2 \alpha_{PR}(T)}
{V^2+2 b N V-b^2 N^2}.
\end{align}
}
Moreover, the critical point is determined by the conditions
\begin{equation}
\left(\frac{\partial P}{\partial V}\right)_{T,N} = 0,
\qquad
\left(\frac{\partial^2 P}{\partial V^2}\right)_{T,N} = 0.
\end{equation}
\begin{figure}[ht!]
\centering

\begin{minipage}{0.498\textwidth}
    \centering
    \includegraphics[width=\linewidth]{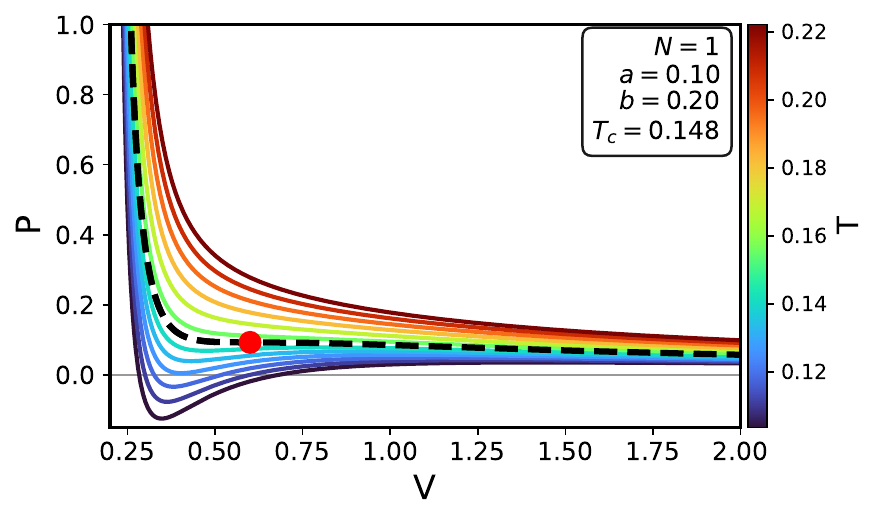}
\end{minipage}
\hfill
\begin{minipage}{0.49\textwidth}
    \centering
    \includegraphics[width=\linewidth]
    {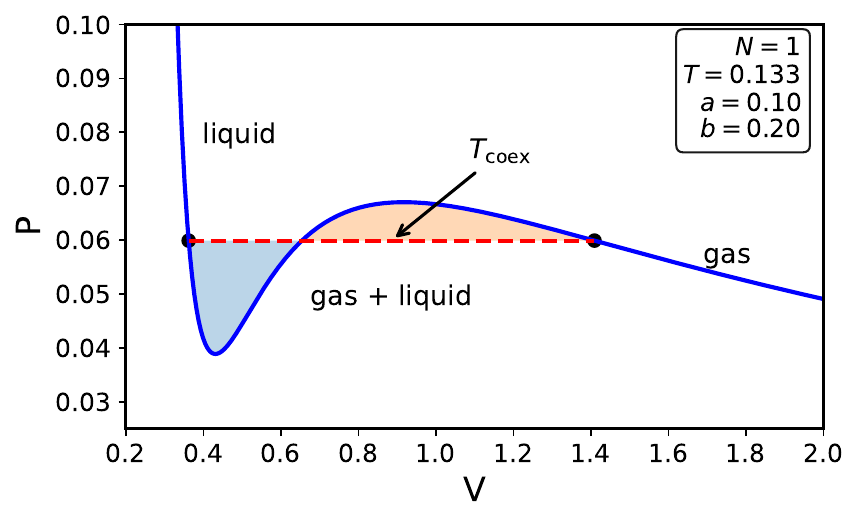}
\end{minipage}
\caption{
(a) $P$--$V$ isotherms of the vdW model for fixed values of $N$, $a$, and $b$, illustrating the evolution of the EoS with temperature $T$.
(b) Maxwell equal-area construction applied to the oscillatory region of the vdW isotherm, replacing the mechanically unstable branch with the coexistence line corresponding to the first-order phase transition.
}
\label{fig:vdw}
\end{figure}
These conditions determine the inflection point of the isotherms in the $P$--$V$ plane at fixed temperature $T$ and particle number $N$, corresponding to the critical point of the system. Solving the above equations yields the critical volume $V_c$, given by
\begin{align}
\textbf{vdW-like: } &\quad
\left( \frac{\partial^3 A}{\partial V^3} \right)_{T,N}
= \frac{2}{V - b N}\left( \frac{\partial^2 A}{\partial V^2} \right)_{T,N},
\\
\textbf{modified vdW: } &\quad
\left( \frac{\partial^4 A}{\partial T\,\partial V^3} \right)_{N}
=
\frac{\left( \frac{\partial^3 A}{\partial T\,\partial V^2} \right)_{N}
\left[2  Nk_BT - (V - b N)^3 \left( \frac{\partial^3 A}{\partial V^3} \right)_{T,N}\right]}
{ Nk_BT (V - b N) - (V - b N)^3 \left( \frac{\partial^2 A}{\partial V^2} \right)_{T,N}}.
\end{align}
Substituting Eqs.~\eqref{eq:vdW}--\eqref{eq:pr} into the previous critical conditions yields the critical volumes
{\small
\begin{align}
&&\textbf{van der Waals: }&\quad V_c=3 b N,\\
&&\textbf{Berthelot: }&\quad V_c=3 b N,\\
&&\textbf{Redlich-Kwong: }&\quad V_c=\left(2^{1/3} - 1\right)^{-1}b N,\\
&&\textbf{Peng-Robinson: }&\quad V_c=\left[1+\left(4+2\sqrt{2}\right)^{1/3}+\left(4-2\sqrt{2}\right)^{1/3}\right]b N =\chi b N.
\end{align}
}
Accordingly, the corresponding critical temperatures and pressures are given by
{\small
\begin{align}
\textbf{van der Waals:}\quad 
& T_c = \frac{8 a}{27 k_B b}, 
&\quad 
& P_c = \frac{a}{27 b^2}, \\
\textbf{Berthelot:}\quad 
& T_c = \sqrt{\frac{8 a}{27 k_B b}}, 
&\quad 
& P_c = \sqrt{\frac{a k_B}{216 b^3}}, \\
\textbf{Redlich--Kwong:}\quad 
& T_c = 3^{2/3}(2^{1/3}-1)^{4/3}
\left(\frac{a}{k_B b}\right)^{2/3}, 
&\quad 
& P_c =
\frac{\left[17 (1 + 2^{4/3}) - 37\,2^{2/3}\right]}{36}
\frac{k_B}{b}
\left(\frac{a}{k_B b}\right)^{2/3}, \\
\textbf{Peng--Robinson:}\quad 
& T_c =
\frac{2(\chi-1)^2(1+\chi)a}
{\left[\chi(2+\chi)-1\right]^2 k_B b},
&\quad 
& P_c =
\frac{\left[\chi(\chi-2)-1\right]a}
{\left[\chi(2+\chi)-1\right]^2 b^2}.
\label{PRcritic}
\end{align}
}
where $\chi$ is a dimensionless constant defined through the Peng--Robinson critical volume. Fig.~\ref{fig:vdw}(a) shows the $P$--$V$ diagram of the prototypical vdW fluid model. Below the critical temperature, the isotherms develop the characteristic oscillatory behavior of the coexistence region. A zoom of this region is presented in Fig.~\ref{fig:vdw}(b), where the unphysical oscillation is replaced by a constant-pressure line through the Maxwell equal-area construction. The order parameter of the liquid--gas phase transition is defined as $\eta \equiv (n_l-n_g)/n_c$, where $n_l$, $n_g$, and $n_c$ are the liquid, gas, and critical densities, respectively. As shown in Fig.~\ref{fig:vdw_coexistence}(a), $\eta$ vanishes at the critical point and remains nonzero for $T<T_c$. Fig.~\ref{fig:vdw_coexistence}(b) displays the reduced phase diagram in the $(P/P_c,,T/T_c)$ plane. The coexistence curve separates the liquid and gas phases and terminates at the critical point, while for $T>T_c$ the system enters the supercritical regime. Similar phase diagrams are obtained for all models considered in this work.

Finally, using the definition of the critical compressibility factor, $Z_c = P_c V_c/ N k_BT_c$, we obtain
\begin{align}
\textbf{van der Waals:}\quad 
& Z_c = \frac{3}{8}  \;\approx\; 0.375,\\[6pt]
\textbf{Berthelot:}\quad 
& Z_c = \frac{3}{8} \;\approx\; 0.375,\\[6pt]
\textbf{Redlich-Kwong:}\quad 
& Z_c = \frac{1}{3} \;\approx\; 0.333,\\[6pt]
\textbf{Peng-Robinson:}\quad 
& Z_c = \frac{\chi (\chi - 2) - 1}{2 (\chi - 1)^2 (1 + \chi)} \;\approx\; 0.307,
\end{align}
showing that $Z_c$ is a universal constant for each model, independent of $a$ and $b$, although its value can differ among the considered EoS models.
%%%%%%%%%%%%%%%%%%%%%%%%%%%%%%%%%%%%%
\subsection{Virial Coefficients and Boyle Temperature}
Real fluids can also be described through the virial expansion, which expresses the compressibility factor $Z$ as a series in the particle density $N/V$
\begin{align}
Z=\frac{PV}{N k_B T}
 = 1 + \frac{N}{V}B(T)
 + \mathcal{O}\!\left(\frac{N^2}{V^2}\right).
\end{align}
Here, $B(T)$ represents the second virial coefficient associated with effective two-body interactions. The temperature at which this coefficient vanishes, defines the Boyle temperature $T_B$. At this point, attractive and repulsive contributions effectively cancel, and the real gas behaves approximately as an ideal gas to leading order. In analogy with the virial expansion, the compressibility factor can be expressed through the function $A$, which encodes the deviations from ideal behavior due to intermolecular interactions. The resulting expressions for $Z$ read
{\small
\begin{align}
\textbf{vdW-like: } &\quad 
Z = 1 + \frac{N\,B(T)}{V}
+ \frac{1}{Nk_B T}
\left(\frac{\partial A}{\partial V}\right)_{T},
\\[6pt]
\textbf{Modified vdW: } &\quad 
Z = 1 + \frac{N\,B(T)}{V}
+ \frac{1}{Nk_B T}
\left(\frac{\partial A}{\partial V}\right)_{T}
+ \frac{1}{Nk_B\!\left(1+\Theta(T)\right)}
\left(\frac{\partial^2 A}{\partial T\,\partial V}\right).
\end{align}
}
Then, the corresponding Boyle temperature is determined from the condition $B(T_B)=0$ and is given by
{\small
\begin{align}
\textbf{vdW-like: } &\quad 
B(T_B)=Nk_B b+\frac{V^2}{NT_B}\left(\frac{\partial A}{\partial V}\right)_T=0,
\quad \Rightarrow\quad 
T_B=-\frac{V^2}{N^2k_B b}\left(\frac{\partial A}{\partial V}\right)_T ,
\\[10pt]
\textbf{Modified vdW: } &\quad 
B(T_B)=Nk_B b+\frac{V^2}{NT_B}\left(\frac{\partial A}{\partial V}\right)_T
+\frac{V^2}{Nk_B\!\left[1+\Theta(T_B)\right]}
\left(\frac{\partial^2 A}{\partial T\,\partial V}\right)=0.
\end{align}
}
\begin{table}[ht!]
\centering
\renewcommand{\arraystretch}{2.5}
\begin{tabular}{l|c|c|c}
\hline
\textbf{Model} & \boldmath{$B(T)$} & \boldmath{$T_{\text{B}}$} & \boldmath{$T_B/T_c$} \\ \hline

\textbf{van der Waals} & 
$b - \dfrac{a}{k_B T}$ & 
$\dfrac{a}{k_B b}$ & 
$\dfrac{27}{8} = 3.375$ \\ \hline

\textbf{Berthelot} & 
$b - \dfrac{a}{k_B T^2}$ & 
$\sqrt{\dfrac{a}{k_B b}}$ & 
$\sqrt{\dfrac{27}{8}} \approx 1.837$ \\ \hline

\textbf{Redlich--Kwong} & 
$b - \dfrac{a}{k_B T^{3/2}}$ & 
$\left( \dfrac{a}{k_B b} \right)^{2/3}$ & 
$\dfrac{1}{3^{2/3}(2^{1/3}-1)^{4/3}} \approx 2.898$ \\ \hline

\textbf{Peng--Robinson} & 
$b-a\frac{\alpha_{PR}(T)}{k_B T}$ & 
$\frac{a(1+m)^2 T_c}{\left(\sqrt{a}\,m + \sqrt{k_B b T_c}\right)^2}$ & 
$\frac{a(1+m)^2}{\left(\sqrt{a}m+\sqrt{k_B b T_c}\right)^2}$ \\ \hline

\end{tabular}
\caption{Analytical expressions for the second virial coefficient $B(T)$, the Boyle temperature $T_B$, and the ratio $T_B/T_c$ for the van der Waals, Berthelot, Redlich--Kwong, and Peng--Robinson models.}
\label{tab:comparison_eos}
\end{table}
The analytical expressions for the virial coefficient $B(T)$, the Boyle temperature $T_B$, and the critical temperature $T_c$ depend on the specific model under consideration. This allows a systematic comparison across different EoS, as summarized in Table~\ref{tab:comparison_eos}.

\section{Geometrothermodynamics and Microstructure} \label{GTD}
Geometrothermodynamics is a geometric framework introduced in
\cite{quevedo2007geometrothermodynamics} that provides a Legendre-invariant
description of thermodynamic systems. Unlike other geometric approaches,
such as thermodynamic geometry
\cite{weinhold1975metric,weinhold1976metric,ruppeiner1979thermodynamics,ruppeiner1981application}, GTD
ensures that the geometric characterization of a thermodynamic system is
independent of the choice of thermodynamic potential
\cite{callen1998thermodynamics}. This invariance allows for a consistent
geometric description of thermodynamics in which phase transitions and
microscopic interactions are encoded in the curvature properties of the
equilibrium manifold
\cite{quevedo2007geometrothermodynamics,quevedo2023unified}. 

%Over the years, GTD has been successfully applied to a wide variety of
%physical systems, including ideal gases and van der Waals fluids
%\cite{quevedo2022geometrothermodynamics,quevedo2011phase}, magnetic
%materials \cite{quevedo2024geometrothermodynamic}, chemical reactions
%\cite{quevedo2014geometric}, cosmological spacetimes
%\cite{romero2026quasi,PhysRevD.86.063508,2019EPJC...79..577B,2014GrCo...20..208Q},
%and black holes in diverse gravitational theories, for example
%\cite{ladino2025phase,ahmed2026shadow,ladino2026probing,Ladino:2024ned}.

To establish the geometric framework employed in the present work, we
briefly summarize the main ingredients of GTD. Legendre invariance is
implemented by introducing an auxiliary $(2n+1)$-dimensional manifold
$\mathcal{T}$, coordinatized by $Z^A=\{\Phi,E^a,I_a\}$, where $n$ is the
number of thermodynamic degrees of freedom and $\Phi$ denotes the
thermodynamic potential \cite{callen1998thermodynamics}. While $E^a$ and
$I_a$ naturally correspond to extensive and intensive variables in
homogeneous systems, this identification becomes subtler for
quasi-homogeneous systems such as black holes
\cite{quevedo2023unified,quevedo2019quasi}. The space $\mathcal{T}$ is endowed with both a local canonical 1-form \(\Theta_G = d\Phi - I_a \, dE^a\), which satisfies the condition of being non-maximally integrable, defining a contact structure,
and a Riemannian metric $G=G_{AB}dZ^A d Z^B$ where $A,B=0,\ldots,2n$. The triplet \((\mathcal{T}, \Theta_G, G)\) defines a Riemannian contact manifold and is referred to as the thermodynamic phase space in the GTD formalism. Currently, there exist three Legendre-invariant metrics on \(\mathcal{T}\), which are given by \cite{quevedo2023unified}
\begin{equation}
    G^{I/II}=\left(d\Phi-I_adE^a\right)^2+(\xi_{ab}E^aI^b)(\chi_{cd}dE^cdI^d), \label{metrics phase1}
\end{equation}
\begin{equation}
    G^{III}=\left(d\Phi-I_adE^a\right)^2+\sum_{a=1}^{n}\xi_{a}(E_aI_a)^{2k+1}dE^adI^a. \label{metric phase2}
\end{equation}
Here, \(\xi_a\) are \(n\) real constants, \(\xi_{ab}\) is a real diagonal \(n \times n\) matrix, and \(k\) is an integer. Moreover, the matrix \(\chi_{cd}\) is defined as \(\chi_{cd} = \delta_{cd} = \text{diag}(1, 1, \ldots, 1)\) for the metric \(G^I\), and as \(\chi_{cd} = \eta_{cd} = \text{diag}(-1, 1, \ldots, 1)\) for \(G^{II}\). In GTD, thermodynamic states are represented as points in an \( n \)-dimensional subspace of \( \mathcal{T} \), known as the equilibrium space \( \mathcal{E} \). This space is defined by a smooth mapping \( \varphi: \mathcal{E} \rightarrow \mathcal{T} \), for which the condition \( \varphi^*(\Theta_G) = 0 \) holds, where $\varphi^\ast$ represents the pullback. As a consequence, the first law of thermodynamics is naturally satisfied on \( \mathcal{E} \), and the coordinates \( Z^A \) become functions of the variables \( E^a \), that is $ Z^A(E^a) = \{ \Phi(E^a), E^a, I_a(E^a) \}$, where \( \Phi = \Phi(E^a) \)  represents the fundamental equation of the thermodynamic system and $I_a=\partial \Phi/\partial E^a $ the dual variables. Additionally, the line element $G = G_{AB}dZ^AdZ^B$ on $\cal{T}$ induces a line element $g = g_{ab}dE^adE^b$ on $\cal{E}$ by means of the pullback, i.e., $\varphi^\star(G) = g$. Then, from Eqs. (\ref{metrics phase1}) -- \eqref{metric phase2}, we obtain
\begin{align}
g^I &= \sum_{a,b,c=1}^{n} \left( \nu_c E^c \frac{\partial \Phi}{\partial E^c} \right) \frac{\partial^2 \Phi}{\partial E^a \partial E^b} \, dE^a \, dE^b,\label{g111} \\
g^{II} &= \sum_{a,b,c,d=1}^{n} \left( \nu_c E^c \frac{\partial \Phi}{\partial E^c} \right) \eta^d_{\;a} \frac{\partial^2 \Phi}{\partial E^b \partial E^d} \, dE^a \, dE^b, \label{g222}\\
g^{III}&=\sum_{a=1}^{n}\nu_a\left(\delta_{ad}E^d\frac{\partial \Phi}{\partial E^a}\right)^{2k+1}\delta^{ab}\frac{\partial^2 \Phi}{\partial E^b \partial E^c} dE^a dE^c. \label{g333}
\end{align}
As in this work we are interested in describing homogeneous thermodynamic systems, we compute the components of the metrics on \( \mathcal{E} \) by choosing the free constants as \( \xi_a = \nu_a \) and \( \xi_{ab} = \text{diag}(\nu_1, \nu_2, \ldots, \nu_n) \), where \( \nu_a \) are the weights that define the scaling properties of the fundamental equation. Moreover, if the Euler relation, \( \Sigma_a \nu_a E^a \partial \Phi / \partial E^a = \nu_\Phi \Phi \), is satisfied, the conformal factor in $g^I$ and $g^{II}$ is replaced by \( \nu_\Phi \Phi \), where \( \nu_\Phi \) is the homogeneity degree of the potential \( \Phi \)~\cite{quevedo2023unified}. Additionally, to ensure that the three metrics describe the same thermodynamic system, we fix $k=0$ in $g^{III}$. With these considerations, the metrics take the form
\begin{align}
    g^I_{ab}&=\nu_\Phi \Phi \delta^c_a\frac{\partial^2 \Phi}{\partial E^b \partial E^c}, \label{g1}\\
    g^{II}_{ab}&=\nu_\Phi \Phi \eta^c_a\frac{\partial^2 \Phi}{\partial E^b \partial E^c}. \label{g2}\\
g^{III}&=\sum_{a=1}^{n}\nu_a\left(\delta_{ad}E^d\frac{\partial \Phi}{\partial E^a}\right)\delta^{ab}\frac{\partial^2 \Phi}{\partial E^b \partial E^c} dE^a dE^c. \label{g3} 
\end{align}
The geometry of $\mathcal{E}$ encodes the thermodynamic properties of
the system, with its curvature providing a geometric measure of the underlying thermodynamic interactions. The GTD metrics $g^{I}$, $g^{II}$, and $g^{III}$ defined in Eqs.~(\ref{g1})--(\ref{g3}) give rise to the corresponding Ricci curvature scalars $\mathcal{R}^{I}$, $\mathcal{R}^{II}$, and $\mathcal{R}^{III}$, respectively. In particular, curvature
singularities are generally associated with phase
transitions, and thermodynamic instabilities \cite{quevedo2007geometrothermodynamics}. Furthermore, according to the interaction hypothesis proposed by
Ruppeiner \cite{ruppeiner1979thermodynamics,ruppeiner1981application,ruppeiner2010thermodynamic},
the sign of the scalar curvature encodes the dominant character of the
effective microscopic interactions: $\mathcal{R}>0$ indicates that
repulsive interactions prevail, whereas $\mathcal{R}<0$ corresponds to
predominantly attractive interactions. The limiting case $\mathcal{R}=0$ is traditionally associated with an ideal-gas-like system, where effective microscopic interactions are absent. However, recent studies have shown that vanishing thermodynamic curvature does not necessarily imply the absence of microscopic interactions \cite{rodrigo2025interacting}, highlighting that the relation between zero curvature and thermodynamic microstructure can be more subtle than the ideal-gas correspondence. Motivated by these considerations, we now investigate the equilibrium space of the vdW model as a prototype system, while a comprehensive analysis of all fluid models is deferred to Section~\ref{universa}.

\subsection{Geometrothermodynamics of the van der Waals Fluid Model}
\label{vdW GTD}
We adopt the entropy representation by choosing the thermodynamic potential
$\Phi \equiv S(U,V,N)$. Since entropy is defined up to an additive constant,
we set the reference entropy $S_{0}=0$ without loss of generality. Combining
Eq.~\eqref{GeneralEntropySimple} and \eqref{eq:vdW}, the fundamental
equation for the vdW model can be expressed as
\begin{align}
S(U,V,N) =  N k_B \left[ \frac{3}{2} \ln \left( \frac{U}{N} + \frac{a N}{V} \right) + \ln \left( \frac{V}{N} - b \right) \right],
\label{funda vdW}
\end{align}
where the vdW parameters $a$ and $b$ are kept explicit throughout the analysis, instead of introducing reduced variables via the law of corresponding states \cite{quevedo2022geometrothermodynamics,quevedo2023geometrothermodynamic}. This choice allows us to interpret the GTD curvature directly in terms of the physical parameters of the fluid. To examine the scaling behavior of the fundamental relation Eq.~\eqref{funda vdW}, we perform a uniform rescaling of the extensive variables by an arbitrary positive parameter $\lambda$
\begin{equation}
U\rightarrow \lambda U, \qquad
V\rightarrow \lambda V, \qquad
N\rightarrow \lambda N.
\end{equation}
Substituting these transformations into Eq.~\eqref{funda vdW}, we readily obtain
\begin{equation}
S(\lambda U,\lambda V,\lambda N)
=
\lambda S(U,V,N),
\end{equation}which shows that the entropy is a homogeneous function of degree one in the extensive variables. Consequently, the Euler identity becomes
\begin{equation}
U\frac{\partial S}{\partial U}
+V\frac{\partial S}{\partial V}
+N\frac{\partial S}{\partial N}
=S,
\end{equation} which is identically satisfied. However, when expressed in terms of the normalized variables $u=U/N$ and $v=V/N$, the entropy per particle $s=S/N$ is no longer a homogeneous function. The same situation arises for the reduced variables introduced through the law of corresponding states \cite{quevedo2022geometrothermodynamics}. In contrast, the fundamental relation Eq.~\eqref{funda vdW}  belongs to the class of homogeneous thermodynamic systems. Consequently, the GTD metrics given in Eqs.~(\ref{g1})--(\ref{g3}) reduce to
\begin{align}
g^{I} =&\, S\ \Bigg[
S_{,UU}\,dU^2
+S_{,VV}\,dV^2
+S_{,NN}\,dN^2
+2S_{,UV}\,dU\,dV
+2S_{,UN}\,dU\,dN
+2S_{,VN}\,dV\,dN
\Bigg],\\[6pt]
g^{II} =&\, S\Bigg[
- S_{,UU}\,dU^2
+ S_{,VV}\,dV^2
+ S_{,NN}\,dN^2
+2S_{,VN}\,dV\,dN
\Bigg],\\[6pt]
g^{III} =&\,
\left(\frac{U}{T}\right)S_{,UU}dU^2
+\left(\frac{PV}{T}\right)S_{,VV}dV^2
-\left(\frac{\mu N}{T}\right)S_{,NN}dN^2 \notag\\
&+S_{,UV}\left[\left(\frac{U}{T}\right)+\left(\frac{PV}{T}\right)\right]dU\,dV
+S_{,UN}\left[\left(\frac{U}{T}\right)-\left(\frac{\mu N}{T}\right)\right]dU\,dN \notag\\
&+S_{,VN}\left[\left(\frac{PV}{T}\right)-\left(\frac{\mu N}{T}\right)\right]dV\,dN;
\end{align}
where the EoS ~\eqref{EoS vdW} has been used to construct the metric $g^{III}$. In this section, we probe only the Ricci scalar associated with $g^{II}$ as a representative example, leaving a comprehensive analysis of all three GTD metrics to Section~\ref{universa}. Therefore, $\mathcal{R}^{II}$ takes the form
\begin{equation}
\mathcal{R}^{II}(U,V,N) =\frac{\mathcal{N}^{II}}{ 3    Nk_B   U^2 V^2 (2S)^3 
\left[
a N^2 \left(3 b^2 N^2 - 6 b N V + 2 V^2\right) - U V^3
\right]^2},
\end{equation}
\begin{figure}[ht!]
\centering

\begin{minipage}{0.32\textwidth}
    \centering
    \includegraphics[width=1.01\linewidth]{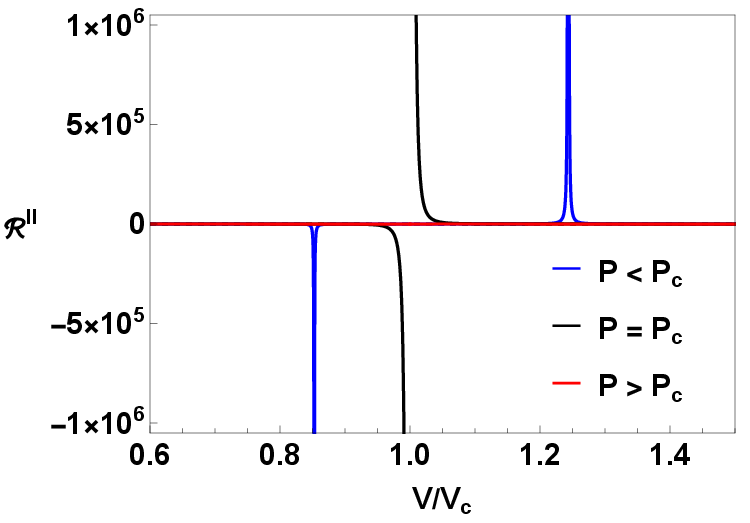}
    (a) $\hspace{0.2cm}a<b,\hspace{0.2cm}T_{B}<1\hspace{1.4cm}$
\end{minipage}
\hfill
\begin{minipage}{0.32\textwidth}
    \centering
    \includegraphics[width=1.01\linewidth]{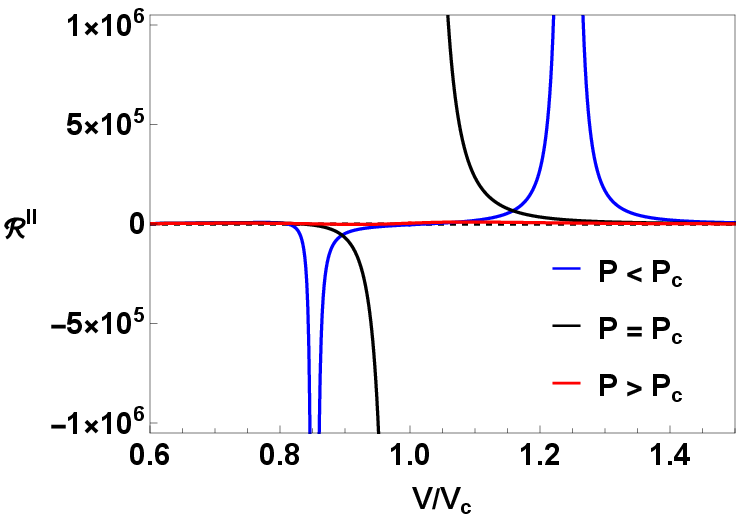}
    (b)$\hspace{0.2cm}a=b,\hspace{0.2cm}T_{B}=1\hspace{1.4cm}$
\end{minipage}
\hfill
\begin{minipage}{0.32\textwidth}
    \centering
\includegraphics[width=1.01\linewidth]{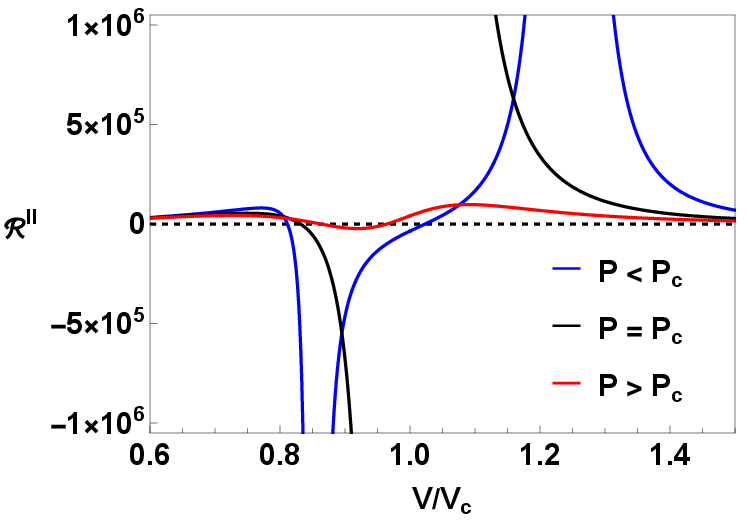}
    (c)$\hspace{0.2cm}a>b,\hspace{0.2cm}T_{B}>1\hspace{1.4cm}$
\end{minipage}
\hfill \\[2em]
\begin{minipage}{0.32\textwidth}
    \centering
    \includegraphics[width=1.01\linewidth]{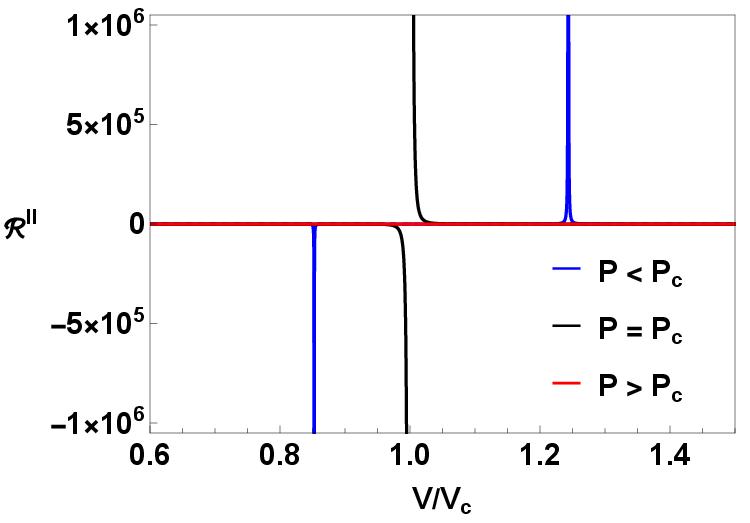}
    (d)$\hspace{0.2cm}N=1\hspace{3.5cm}$
\end{minipage}
\hfill
\begin{minipage}{0.32\textwidth}
    \centering
    \includegraphics[width=1.01\linewidth]{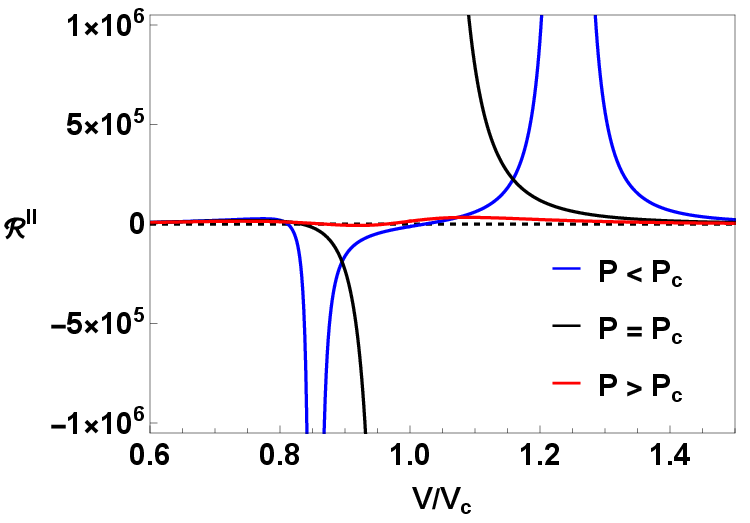}
    (e)$\hspace{0.2cm}N=15\hspace{3.3cm}$
\end{minipage}
\hfill
\begin{minipage}{0.32\textwidth}
    \centering
    \includegraphics[width=1.01\linewidth]{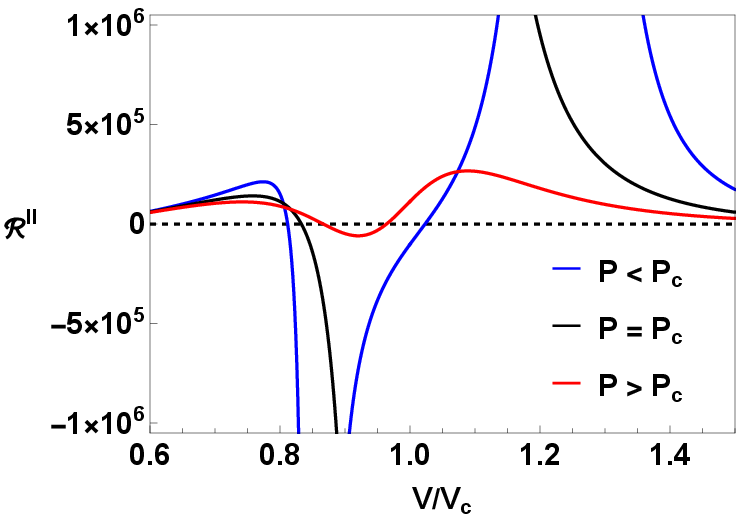}
    (f)$\hspace{0.2cm}N=30\hspace{3.3cm}$
\end{minipage}
\caption{GTD scalar curvature $\mathcal{R}^{II}$ versus reduced volume $V/V_c$ for the vdW model under different isobaric regimes relative to $P_c$ ($k_B=1$, $S_0=0$). Panels (a)--(c) show the effect of $a$ and $b$ at fixed $N=10$, while (d)--(f) illustrate the dependence on $N$ for $a=b=15$. The curvature is normalized as $\mathcal{R}^{II}\to \mathcal{R}^{II}S^3U^2$.}
\label{fig:vdwScalar2Figures}
\end{figure}
where $\mathcal{N}^{II}$ is a polynomial whose explicit expression is too lengthy to be displayed here. In Fig.~\ref{fig:vdwScalar2Figures}, we illustrate the behavior of $\mathcal{R}^{II}$ for different parameter values. In particular, it is observed that the scalar curvature exhibits a single divergence for $P<P_c$, two divergences at $P=P_c$, and remains regular for $P>P_c$. The effect of varying the ratio $a/b$ is shown in Figs.~\ref{fig:vdwScalar2Figures}(a)--(c), while the influence of $N$ is presented in Figs.~\ref{fig:vdwScalar2Figures}(d)--(e).
\begin{figure}[ht!]
\centering

\begin{minipage}{0.48\textwidth}
    \centering
    \includegraphics[width=\linewidth]{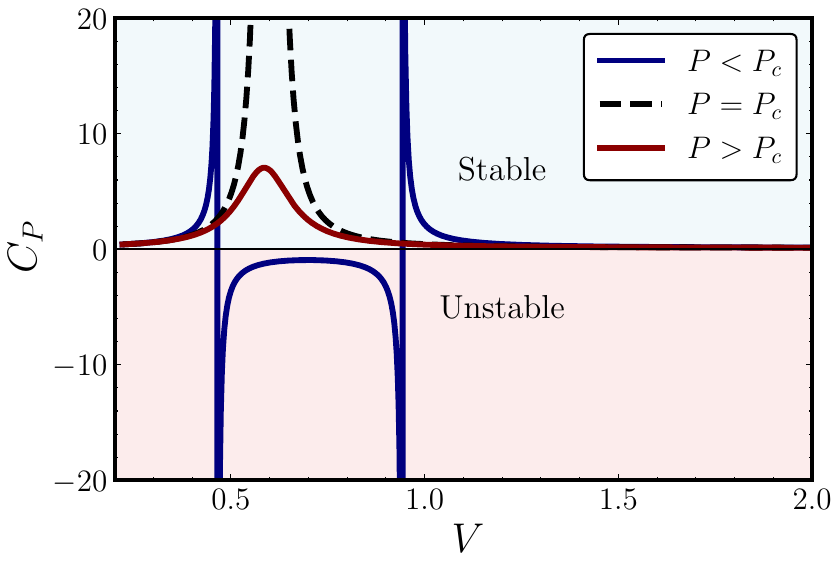}

\end{minipage}

\caption{
(a) Heat capacity $C_P$ of the vdW model as a function of the volume $V$ for $P<P_c$, $P=P_c$, and $P>P_c$, with fixed parameters $a=0.1$, $b=0.2$, and $N=k_B=1$. The heat capacity exhibits the same divergence structure as the scalar curvature, with singularities occurring at the corresponding phase transition points.
}

\label{fig:vdw_response}
\end{figure}
The divergence structure of the scalar curvature coincides with that of the heat capacity at constant pressure
\begin{align}
C_{P}\equiv T\left(\frac{\partial S}{\partial T}\right)_{P}
=\frac{k_B N}{2}
\left[
\frac{
a N^2 (3 b N - 4 V)(3 b N - 2 V) - U V^3
}{
U V^3 + a N^2 V^2 - 3 a N^2 (V - bN)^2
}
\right],
\label{eq:cp}
\end{align}
whose singularities occur at the same thermodynamic critical points. Thus, it is clear that GTD accurately reproduces the local phase structure of the system. This result follows from the one-to-one correspondence between the divergences of thermodynamic response functions and the singularities of the GTD scalar curvature in quasi-homogeneous systems\footnote{Homogeneous systems constitute a particular class of quasi-homogeneous systems, corresponding to equal scaling weights.}; see Refs.~\cite{Ladino:2024ned,ladino2025phase,ladino2026probing} for further details. 

\begin{figure}[ht!]
\centering

\begin{minipage}{0.48\textwidth}
    \centering
    \includegraphics[width=1.01\linewidth]{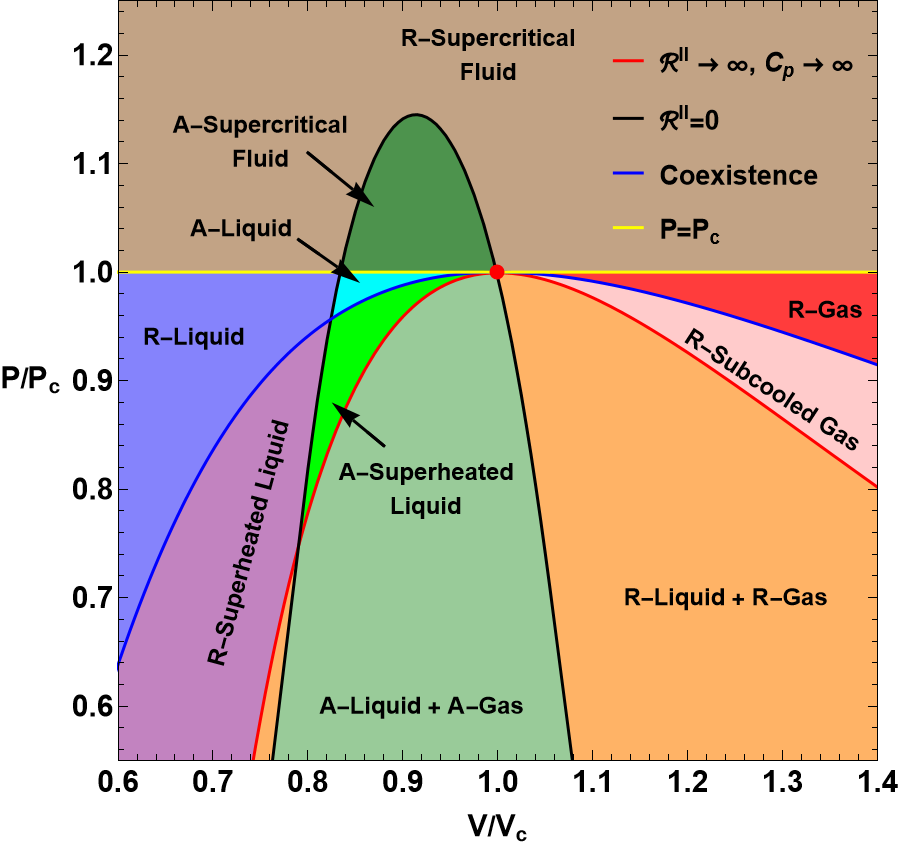}
    (a)$\hspace{7cm}$
\end{minipage}
\hfill
\begin{minipage}{0.48\textwidth}
    \centering
    \includegraphics[width=1.01\linewidth]{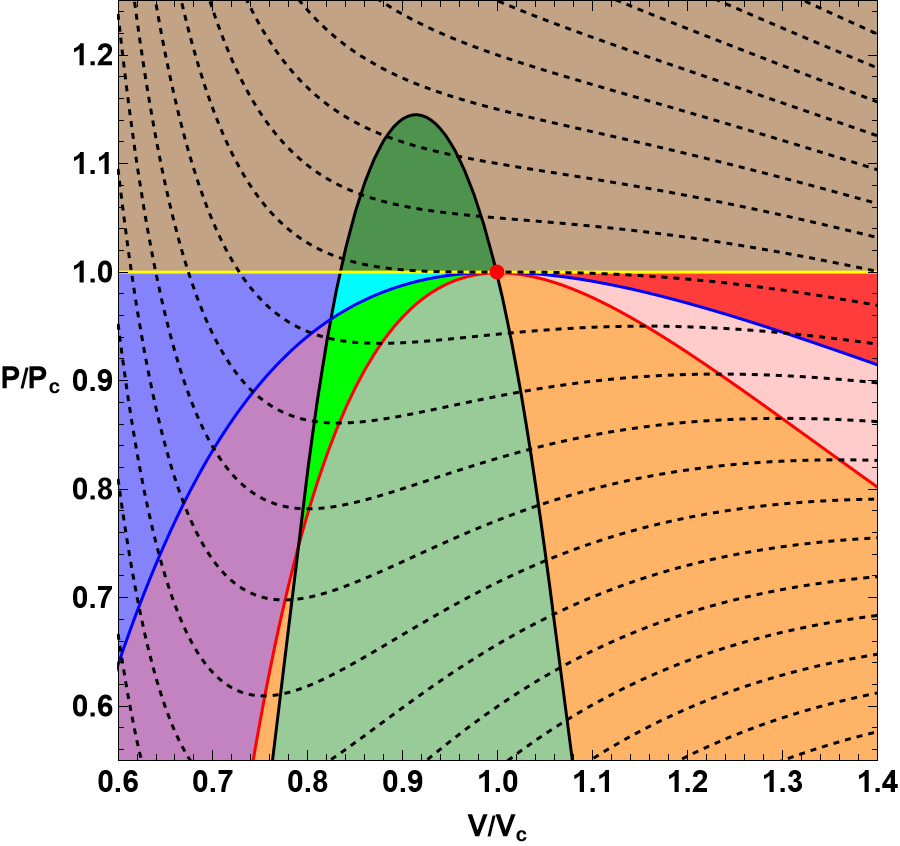}
    (b)$\hspace{7cm}$
\end{minipage}

\caption{Microscopic phase diagrams of the vdW model. Stability and metastability regions are shown, with attractive (A$-$) and repulsive (R$-$) interactions determined by the sign of $\mathcal{R}^{II}$. The binodal (blue), spinodal (red), and zero-curvature (black) curves are displayed; the red dot marks the critical point ($k_B=1$, $S_0=0$, $a=b=20$, $N=5$). Panel (a) labels the phases, while (b) shows isotherms for $T\in[0.8T_c,1.1T_c]$.}
\label{fig:vdwPhaseDiagrams}
\end{figure}

\begin{figure}[ht!]
\centering
\begin{minipage}{0.325\textwidth}
    \centering
    \includegraphics[width=\linewidth]
    {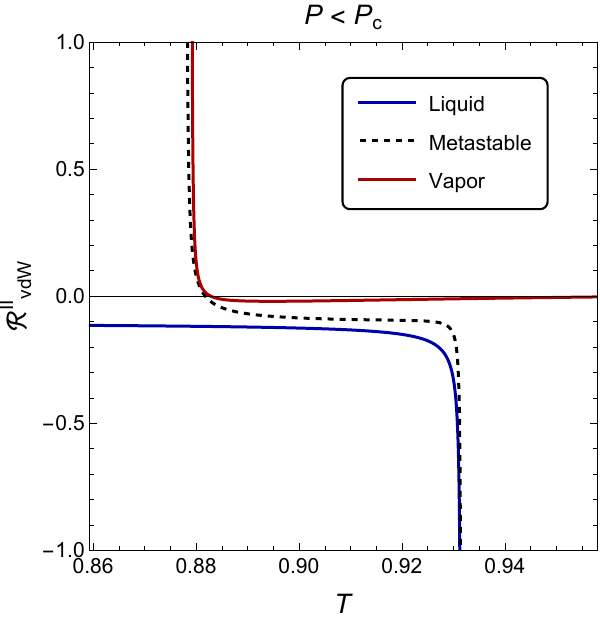}
    
    (a)
\end{minipage}
\hfill
\begin{minipage}{0.315\textwidth}
    \centering
    \includegraphics[width=\linewidth]
    {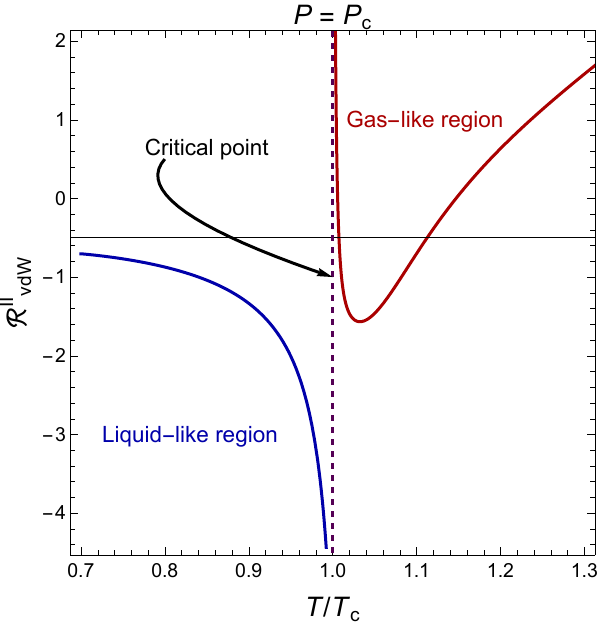}
    
    (b)
\end{minipage}
\hfill
\begin{minipage}{0.325\textwidth}
    \centering
    \includegraphics[width=\linewidth]
    {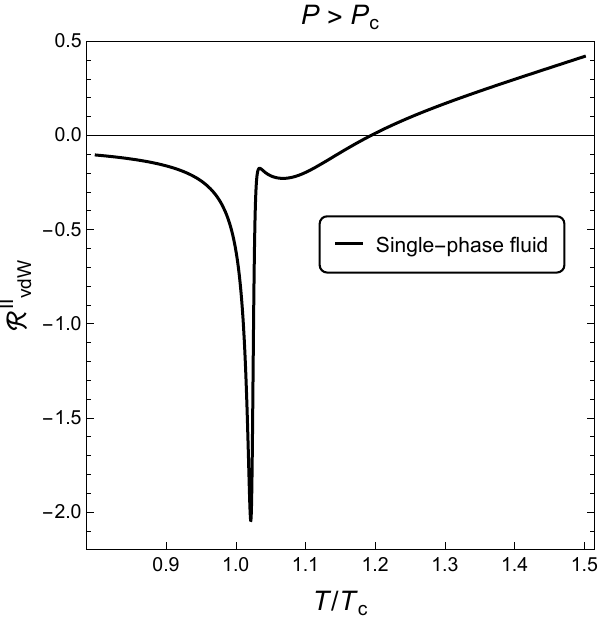}
    
    (c)
\end{minipage}
\caption{
Normalized Ricci scalar $F^3 \mathcal{R^{II}}$ as a function of $T$ for a vdW model with parameters $a=1$, $b=0.3$, and $N=k_B=1$. Panels correspond to (a) $P=0.7P_c$, (b) $P=P_c$, and (c) $P=1.1P_c$. }

\label{fig:vdw microstructure}
\end{figure}
Additionally, Fig.~\ref{fig:vdwPhaseDiagrams} illustrates the phase structure of the vdW model in the equilibrium space. Fig.~\ref{fig:vdwPhaseDiagrams}(a) identifies the different thermodynamic phases and coexistence regions, alongside the attractive ($\mathcal{R}^{II}<0$) and repulsive ($\mathcal{R}^{II}>0$) interaction domains. Furthermore, Fig.~\ref{fig:vdwPhaseDiagrams}(b) displays the corresponding isotherms in the equilibrium manifold, demonstrating how the phase boundaries and critical behavior are encoded within the geometric structure of the system. Notably, the corresponding phase structures and isotherms for the Berthelot, Redlich–Kwong, and Peng–Robinson models are nearly identical to those shown for the vdW model in Fig.~\ref{fig:vdwPhaseDiagrams}.
Finally, we probe the thermodynamic microstructure of the vdW fluid through the GTD scalar $\mathcal{R}^{II}$ obtained from the Helmholtz free energy, Eq.~\eqref{Free vdW}. The resulting normalized quantity $F^3\mathcal{R}^{II}$ is displayed in Fig.~\ref{fig:vdw microstructure}. For the subcritical isobar, $P<P_c$ (Fig.~\ref{fig:vdw microstructure}(a)), the EoS admits up to three solutions for $V(P,T)$, corresponding to the liquid, metastable (unstable), and vapor phases, in agreement with the phase structure shown in Fig.~\ref{fig:vdw}. At low temperatures only the liquid phase is present, whereas at intermediate temperatures all three branches coexist. As the temperature increases further, only the vapor phase survives. The metastable phase terminates at two spinodal temperatures, where it merges with the liquid and vapor phases, respectively. An important feature revealed by Fig.~\ref{fig:vdw microstructure}(a) is that for the subritical isobar, the vapor phase generally exhibits a smaller magnitude of $|\mathcal{R}^{II}|$ than the liquid phase, indicating weaker effective intermolecular interactions and a behavior closer to that of an ideal gas. At the critical pressure, $P=P_c$ (Fig.~\ref{fig:vdw microstructure}(b)), the liquid and vapor branches continuously merge at the critical temperature, signaling the onset of criticality. And for $P>P_c$ the scalar is regular everywhere (Fig.~\ref{fig:vdw microstructure}(c)), indicating  only a single single fluid phase remains. These results closely resemble the behavior previously reported for AdS black holes, where the vapor and liquid phases are naturally identified with the large- and small-black-hole branches, respectively. In both systems, the low-density (large) phase is characterized by weaker microscopic interactions, while the high-density (small) phase displays a stronger correlation structure \cite{Ladino:2024ned,ladino2025phase}.

\section{Geometric Universality Near Criticality}
\label{universa}
\begin{figure}[ht!]
\centering
\begin{minipage}{0.3\textwidth}
    \centering
    \includegraphics[width=\linewidth, trim=2 2 1 2, clip]{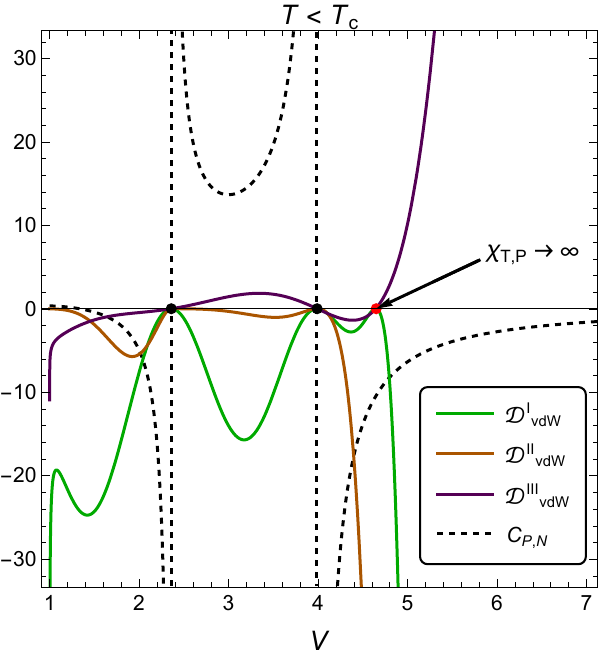}
    (a)
\end{minipage}\hfill
\begin{minipage}{0.3\textwidth}
    \centering
    \includegraphics[width=\linewidth, trim=2 2 1 2, clip]{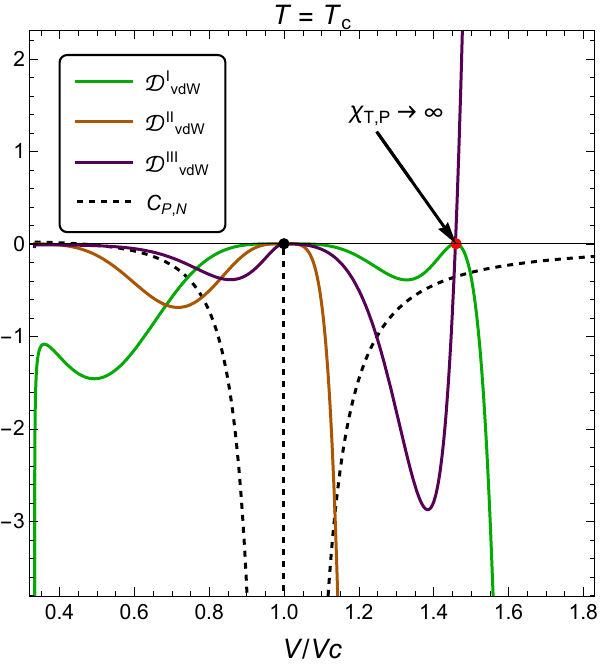}
    (b)
\end{minipage}\hfill
\begin{minipage}{0.3\textwidth}
    \centering
    \includegraphics[width=\linewidth, trim=2 2 1 2, clip]{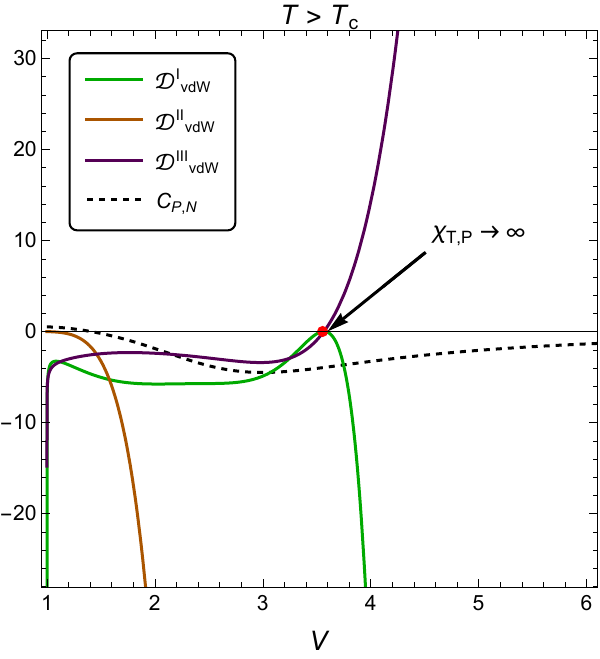}
    (c)
\end{minipage}\hfill\\[0.5em]
\begin{minipage}{0.3\textwidth}
    \centering
    \includegraphics[width=\linewidth, trim=2 2 1 2, clip]{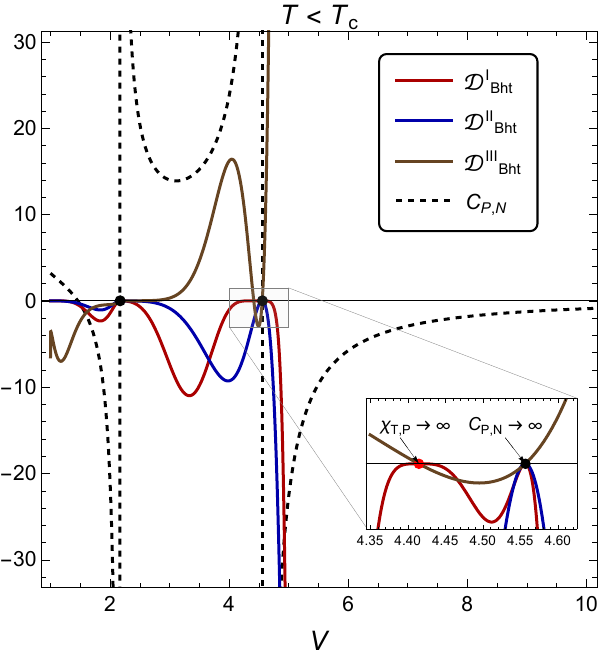}
    (d)
\end{minipage}\hfill
\begin{minipage}{0.3\textwidth}
    \centering
    \includegraphics[width=\linewidth, trim=2 2 1 2, clip]{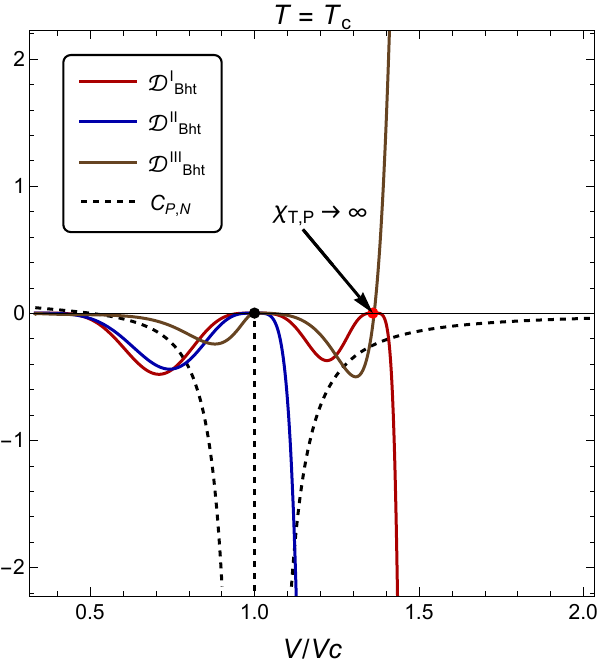}
    (e)
\end{minipage}\hfill
\begin{minipage}{0.3\textwidth}
    \centering
    \includegraphics[width=\linewidth, trim=2 2 1 2, clip]{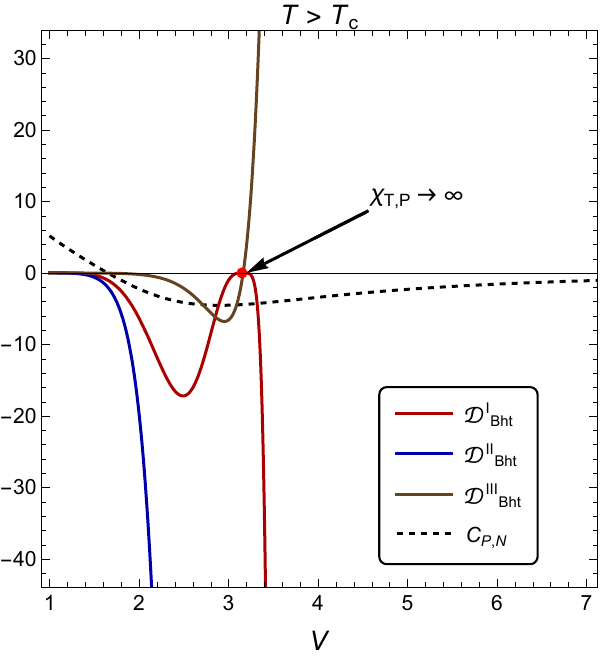}
    (g)
\end{minipage}\hfill\\[0.5em]
\begin{minipage}{0.3\textwidth}
    \centering
    \includegraphics[width=\linewidth, trim=2 2 1 2, clip]{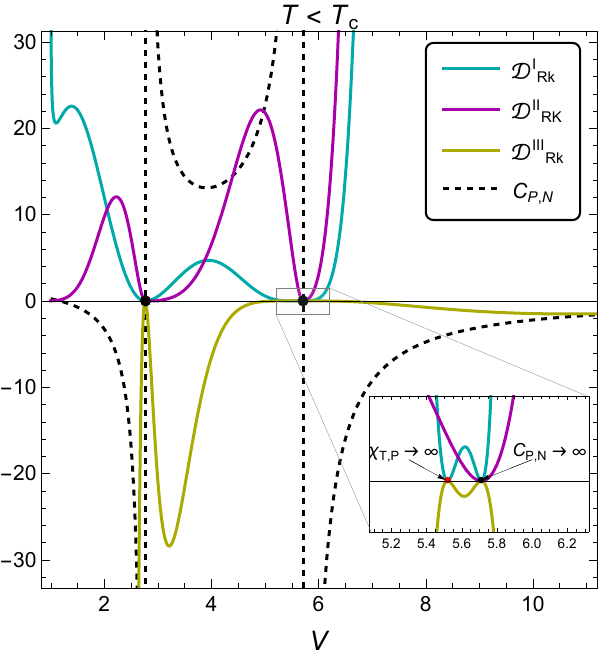}
    (h)
\end{minipage}\hfill
\begin{minipage}{0.3\textwidth}
    \centering
    \includegraphics[width=\linewidth, trim=2 2 1 2, clip]{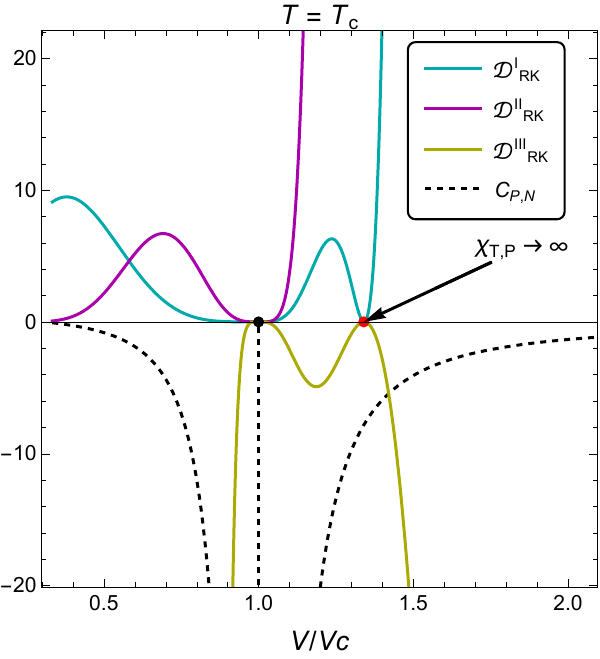}
    (i)
\end{minipage}\hfill
\begin{minipage}{0.3\textwidth}
    \centering
    \includegraphics[width=\linewidth, trim=2 2 1 2, clip]{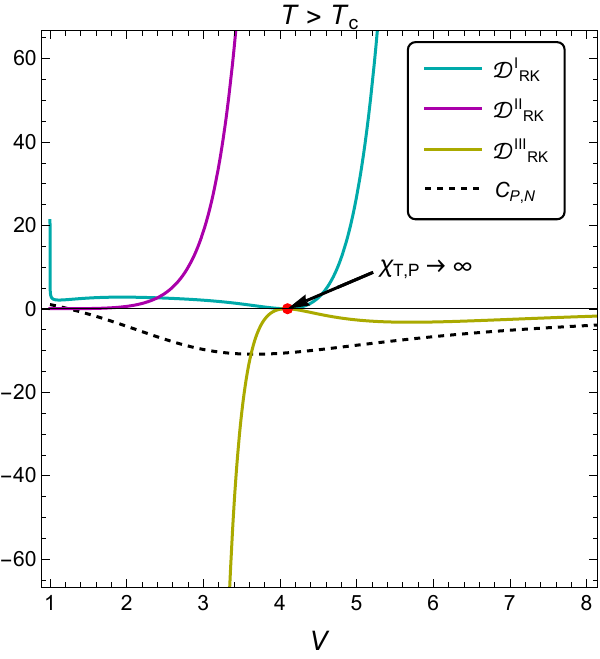}
    (j)
\end{minipage}

\caption{
GTD scalars for $a=b=N=1$, constructed from the Helmholtz free energy at 
$T=0.95T_c$, $T=T_c$, and $T=1.2T_c$. 
Each panel shows the normalized denominator $\mathcal{D}/F^{3}$ associated with 
$\mathcal{R}^{I}$, $\mathcal{R}^{II}$, and $\mathcal{R}^{III}$, together with the corresponding heat capacity. 
Black dots denote the zeros of the denominator that coincide with the divergence points of $C_{P,N}$, whereas red dots indicate the zeros associated with divergences of $\chi_{T,P}$. 
Panels (a)--(c) correspond to the vdW model, panels (d)--(f) to the Bht model, and panels (g)--(i) to the R--K model. 
All curves have been rescaled for visualization purposes.
}
\label{fig:Scalars_Free energy}
\end{figure}

We now analyze the behavior of the GTD scalar curvature in the vicinity of the critical point. In this regime, thermodynamic systems exhibit universal features that are largely independent of the microscopic details of the underlying model. In particular, the curvature develops a characteristic divergence of the form $\mathcal{R} \sim |\tau|^{-\zeta}$ \cite{romero2026quasi,ladino2026probing}. Along the isochoric path $V = V_c$, the curvature admits the expansion
\begin{equation}
\mathcal{R}(\tau) =
\frac{A_c}{|\tau|^\zeta}
+ \mathcal{R}_0 
+ \mathcal{O}(\tau), \qquad 
\tau \equiv \frac{T - T_c}{T_c}, \label{Critical amp}
\end{equation}
where $\mathcal{R}_0 $ is a constant. This expression defines the critical amplitude $A_c$ and the critical exponent $\zeta$. In this section, to compute the GTD curvature, we employ the Helmholtz free energy $F(V,T,N)$, since temperature is the natural control parameter in our analysis. Although the GTD curvature can be equivalently constructed in the entropy representation, as illustrated for the vdW model in Section~\ref{vdW GTD}, the Helmholtz potential is more convenient for studying temperature-driven critical phenomena. Nevertheless, as shown in Ref.~\cite{ladino2026probing}, criticality can be consistently characterized using arbitrary thermodynamic potentials. Using the general expression for $F(V,T,N)$ derived within the unified entropic framework in Eq.~\eqref{Free energy generalized}, we obtain the explicit Helmholtz free energies\footnote{For simplicity, the Peng--Robinson model is omitted from the subsequent GTD analysis due to its qualitatively similar critical behavior.} by substituting the corresponding functions $A(V,N)$ and $\Theta(T)$ from Eqs.~\eqref{eq:vdW}--\eqref{eq:rk} and setting $S_0=0$. These expressions satisfy the Maxwell relations and the first law Eq.~\eqref{first law}, and are given by
\begin{align}
F_{\text{vdW}} &=
-\frac{N}{2V}\left[
2 a N
+ k_B T V \,
\ln\!\left(
\frac{27 (k_B T)^3 (V - b N)^2}{8 e^3 N^2}
\right)
\right],\label{Free vdW} \\[1.2em] 
F_{\text{Bth}} &= 
-\frac{N}{2 T V} \left[
2 a N 
+ k_B T^2 V \ln\!\left(
\frac{27 (k_B T)^3 (V - b N)^2}{8 e^3 N^2}
\right)
\right], \label{Free Bht} \\[1.2em]
F_{\text{R-K}} &=
\frac{N}{2 b \sqrt{T}} \Bigg\{
b\,k_B\,T^{3/2}
\Bigg[
3
+ \ln\!\left(
\frac{8 N^2}{27 (k_B T)^3 (V - b N)^2}
\right)
\Bigg]
+ 2 a \ln\!\left( \frac{V}{V + b N} \right)
\Bigg\}. \label{Free RK}
\end{align}
Notice that all free energies satisfy the Euler identity
\begin{equation}
\nu_T \left(\frac{\partial F}{\partial T}\right) T
+ \nu_V \left(\frac{\partial F}{\partial V}\right) V
+ \nu_N \left(\frac{\partial F}{\partial N}\right) N = F \, ,
\end{equation}
with weights $\nu_T = 0$ and $\nu_V = \nu_N = 1$. Therefore, the free energy is a homogeneous function of degree one, being extensive in $(V,N)$ while $T$ remains an intensive variable. We then compute the three-dimensional GTD metrics given in Eqs.~\eqref{g111}--\eqref{g333}, using the Helmholtz free energy, Eqs.~\eqref{Free vdW}--\eqref{Free RK}, together with the corresponding scalar curvatures for each model, as shown in Fig.~\ref{fig:Scalars_Free energy}. Furthermore, the resulting critical quantities are summarized in Table~\ref{tab:Ac_models}. The numerators of the scalar curvatures are too complicated to be written explicitly. However, the denominators take the simple form \cite{quevedo2023unified,Ladino:2024ned}
\begin{align}
\mathcal{D}^{I} 
&= 2 F^{3} \Big[
-2 F_{,TN} F_{,VN} F_{,TV}
+ F_{,NN} F_{,TV}^{2}
+ F_{,TN}^{2} F_{,VV}
+ F_{,TT} \left( F_{,VN}^{2} - F_{,NN} F_{,VV} \right)
\Big]^{2}, \\[0.8em]
\mathcal{D}^{II} 
&= 2 F^{3}
\left(
F_{,TN}^{2} - F_{,NN}\, F_{,TT}
\right)^{2}
F_{,VV}^{2}, \\[0.8em]
\mathcal{D}^{III}
&=
NV\,F_{,N}\,F_{,V}
\Big[
V\,F_{,V}\,F_{,TV}
\left(
 F_{,NN}\,F_{,TV}- F_{,TN}\,F_{,VN}
\right)
\nonumber\\
&\qquad\qquad\qquad
+
N\,F_{,N}\,F_{,TN}
\left(
 F_{,TN}\,F_{,VV}- F_{,VN}\,F_{,TV}
\right)
\Big]^{2}.
\end{align}
Moreover, the Nambu bracket formalism enables us to express the denominator of the scalar curvature in a straightforward manner in terms of the generalized response functions of the thermodynamic system in a 3-dimensional coordinate space  (see \cite{romero2024extended1, Ladino:2024ned}).
\begin{align}
\mathcal{D}^{I} 
&= 
2 F^{3}
\left(
\frac{C_{P,\mu}}
{ V\times T\times N\times \chi_{T,P}\times \kappa_{T,N}}
\right)^{2}, 
\\[0.8em]
\mathcal{D}^{II} 
&= 
2 F^{3}
\left(
\frac{C_{V,N}}
{V\times T\times N\times  \chi_{S,V}\times \kappa_{T,N}}
\right)^{2},
\\[0.8em]
\mathcal{D}^{III}
&=
-\mu PVN
\left(
\frac{
 \alpha^{(N)}_{P,\mu}\alpha^{(\mu)}_{V,N}
- \alpha^{(V)}_{P,\mu}\alpha^{(P)}_{V,N}
}{
V\times N\times\,\chi_{T,P}\times\,\kappa_{T,N}
}
\right)^2.
\end{align}
The relevant thermodynamic response functions are defined as
\begin{align}
\chi_{Y,Z}
&=
\frac{1}{N}\left(
\frac{\partial N}{\partial \mu}
\right)_{Y,Z},
\qquad
\kappa_{Y,Z}
=
-\frac{1}{V}
\left(
\frac{\partial V}{\partial P}
\right)_{Y,Z},
\\[0.8em]
C_{Y,Z}
&=
T
\left(
\frac{\partial S}{\partial T}
\right)_{Y,Z},
\qquad
\alpha^{(X)}_{Y,Z}
=
X\left(
\frac{\partial X}{\partial T}
\right)_{Y,Z},
\qquad
X\in\{V,N,\mu,P\};
\end{align}
with $\chi_{Y,Z}$ denoting the chemical susceptibility, $\kappa_{Y,Z}$ the mechanical compressibility, $C_{Y,Z}$ the heat capacity, and $\alpha^{(X)}_{Y,Z}$ the generealized thermal response coefficient associated with the thermodynamic variable $X$, all evaluated while keeping the thermodynamic parameters $(Y,Z)$ fixed \cite{romero2024extended1}. In addition, through the thermodynamic identity \cite{callen1998thermodynamics}
\begin{equation}
    C_{P,N}=C_{V,N}-TV\kappa_{T,N}\left(\frac{\partial P}{\partial T}\right)^{2}_{V,N},\label{Iden_1}
\end{equation}
it is clear that the divergences of $\kappa_{T,N}$ coincide with those of $C_{P,N}$. In contrast, $\chi_{S,V}$, $C_{P,\mu}$, and $\alpha^{(X)}_{Y,Z}$ remain regular throughout the physical domain and therefore do not introduce additional singularities. However, $\mathcal{D}^{I}$ and $\mathcal{D}^{III}$ exhibit an additional zero associated with the divergences of the chemical susceptibility $\chi_{T,P}$, as illustrated in Fig.~\ref{fig:Scalars_Free energy}. It is worth emphasizing that all three metrics consistently reproduce the divergences of $\kappa_{T,N}$ and $C_{P,N}$, thereby yielding an identical phase structure regardless of the thermodynamic metric employed, as illustrated in Fig.~\ref{fig:Scalars_Free energy}. This behavior contrasts with what is observed in quasi-homogeneous thermodynamic systems such as black holes, where the metric $g^{III}$ fails to accurately capture the underlying phase structure \cite{ladino2026probing}. Clarifying the origin of this discrepancy, as well as its possible physical and geometrical implications, remains an important open question that deserves further investigation.

\begin{figure}[ht!]
\centering

\begin{minipage}{0.3\textwidth}
    \centering
    \includegraphics[width=\linewidth]{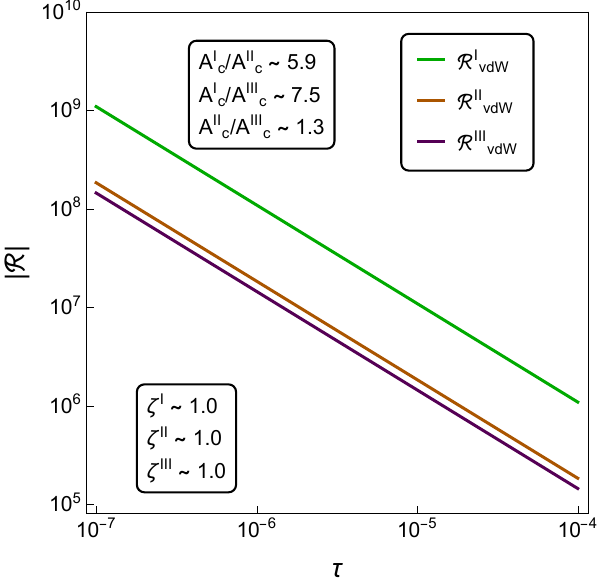}
    (a)
\end{minipage}\hfill
\begin{minipage}{0.3\textwidth}
    \centering
    \includegraphics[width=\linewidth]{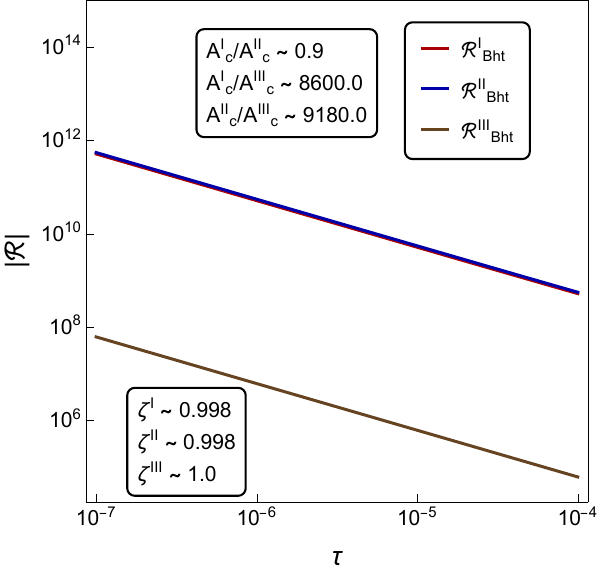}
    (b)
\end{minipage}\hfill
\begin{minipage}{0.3\textwidth}
    \centering
    \includegraphics[width=\linewidth]{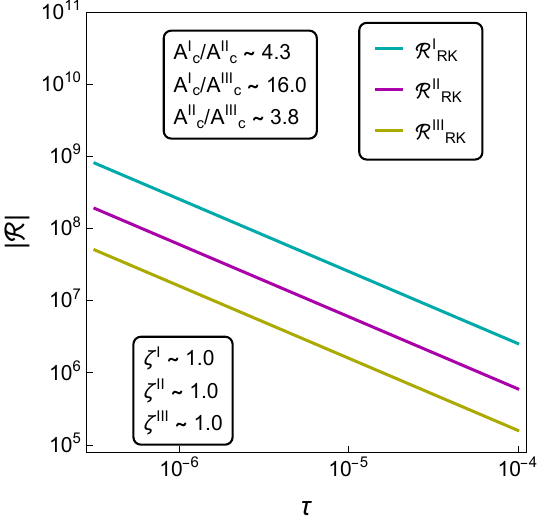}
    (c)
\end{minipage}

\caption{
Log--log plots of the absolute value of the GTD scalar curvature near the critical point as a function of the reduced temperature $\tau$, evaluated at $V = V_c$. A clear power-law divergence $|\mathcal{R}| \sim \tau^{-1}$ is observed, with the slope encoding the critical exponent. Panels (a)--(c) correspond to the different fluid models. All quantities are computed for fixed parameters $a = b = N = 1$.
}
\label{fig:R_tau_loglog}

\end{figure}
\setlength{\tabcolsep}{9pt}
\begin{table}[ht!]
\centering
\large
\begin{tabular}{c c c}
\hline 
Model  & $Q^I_{\;II}\equiv A_c^{I}/A_c^{II}$ & $\zeta$ \\
\hline \\ [0.2em]

van der Waals &
$\frac{\,P_3(x,u)\,P_1(x,u)}{{\ln\!\left(b^2r^3\right)}^2\,P_2(x,u)}$
& $1$ \\ [2em]

Berthelot &
$\frac{\,P_4(x,u)\,P_1(x,u)}{{\ln\!\left(b^8 r^{12}\right)}^4\,P_2(x,u)}$
& $1$ \\ [2em]

Redlich--Kwong &
$\frac{\,P_3(x,u)\,P_1(x,u) P_3(x,u)}{P_7(x,u)}$
& $1$ \\

\hline
\end{tabular}
\caption{
Critical ratios $Q^i_{\;j}\equiv A_c^{i}/A_c^{j}$ in GTD and the universal exponent $\zeta$ in real gas models. 
We define $r = (a/b)^{\delta}$ with $\delta = 1,\,1/2,\,3/2$ for vdW, Berthelot, and Redlich--Kwong, respectively, and $x = \ln b$, $u = \ln(br)$. 
Here, $P_d(x,u)$ denotes a polynomial of degree $d$.
}
\label{tab:Ac_models}
\end{table}

\begin{figure}[ht!]
\centering

\begin{minipage}{0.32\textwidth}
    \centering
    \includegraphics[width=\linewidth]
    {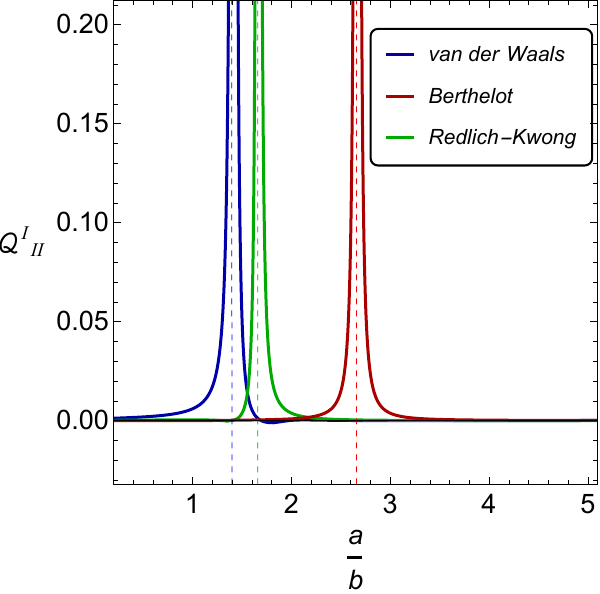}
    
    (a)
\end{minipage}
\hfill
\begin{minipage}{0.32\textwidth}
    \centering
    \includegraphics[width=\linewidth]
    {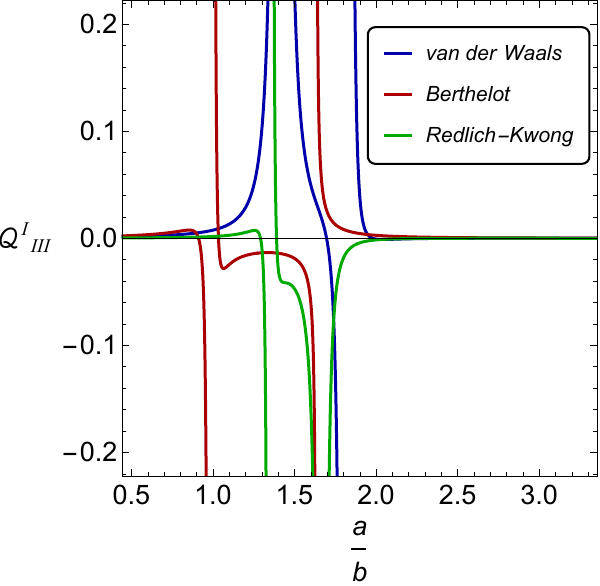}
    
    (b)
\end{minipage}
\hfill
\begin{minipage}{0.33\textwidth}
    \centering
    \includegraphics[width=\linewidth]
    {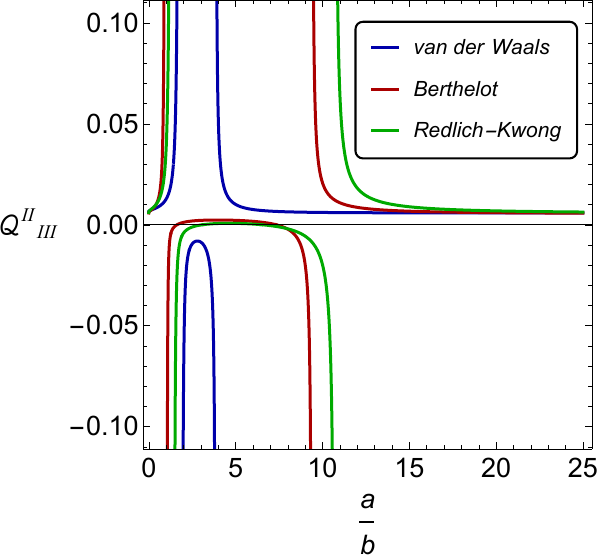}
    
    (c)
\end{minipage}

\caption{
Critical ratios $Q^i_{\;j}$ as functions of the vdW Boyle scale, $a/b$, for all fluid models. (a) $Q^I_{\;II}$, which exhibits the simplest behavior, characterized by a single divergence. (b) and (c) display qualitatively similar behavior; however, an additional divergence appears in both $Q^I_{\;III}$ and $Q^{II}_{\;III}$ for all models. In all panels, the parameters are fixed to $a=b=k_B=1$. Plots have been rescaled for clarity.
}
\label{fig:Qtodos}
\end{figure}

\begin{figure}[ht!]
\centering
\begin{minipage}{0.325\textwidth}
    \centering
    \includegraphics[width=\linewidth]
    {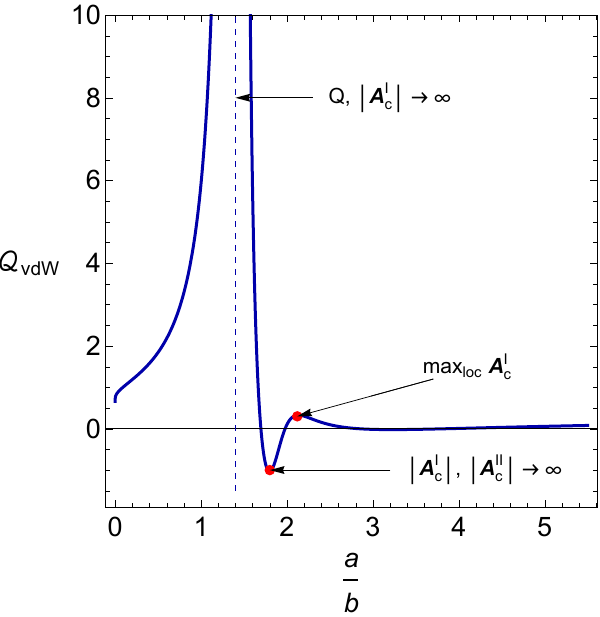}
    
    (a)
\end{minipage}
\hfill
\begin{minipage}{0.315\textwidth}
    \centering
    \includegraphics[width=\linewidth]
    {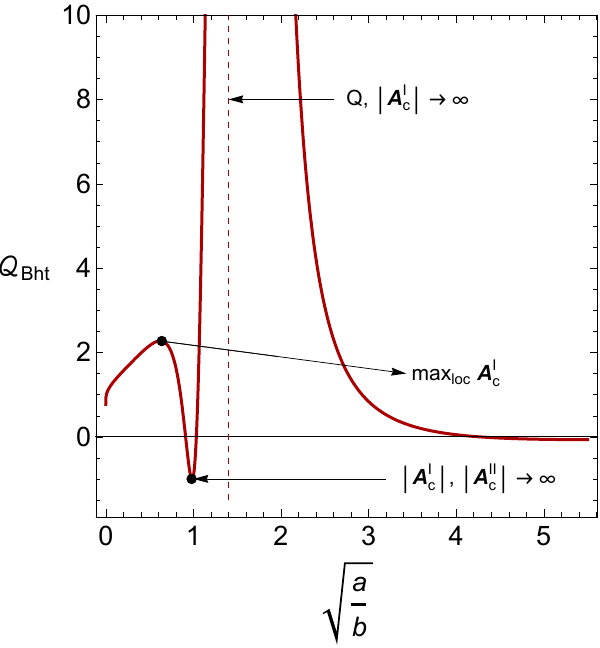}
    
    (b)
\end{minipage}
\hfill
\begin{minipage}{0.325\textwidth}
    \centering
    \includegraphics[width=\linewidth]
    {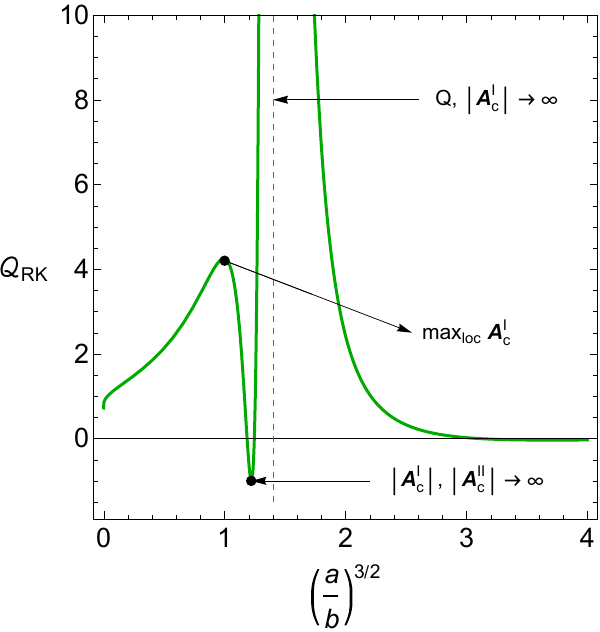}
    
    (c)
\end{minipage}
\caption{Behavior of $Q^{I}_{\;II}$ for (a) the vdW, (b) Berthelot, and (c) Redlich--Kwong models, expressed in terms of their respective Boyle scales. Despite their distinct EoS and microscopic descriptions, all models exhibit the same emergent universal behavior. In all panels, the parameters are fixed to $a=b=k_B=1$.}
\label{fig:Q}
\end{figure}

In Fig.~\ref{fig:R_tau_loglog} and Table~\ref{tab:Ac_models}, we present the  behavior of the GTD scalar curvature in the vicinity of the critical point. The corresponding best-fit results for the critical exponent and the ratio of amplitudes are also displayed. All models exhibit the same critical exponent $\zeta \sim 1$, independently of the GTD metric employed, indicating a universal mean-field behavior. This value differs from that obtained in Ruppeiner geometry \cite{may2013thermodynamic}, where $\zeta = 2$ . Notably, $\zeta = 1$ coincides with the critical scaling of  $\kappa_{T,N}$ and $C_{P,N}$, indicating that the GTD curvature encodes the same critical behavior as standard thermodynamic response functions. This scaling also appears in black hole and cosmological systems \cite{ladino2026probing,romero2026quasi}, highlighting the robustness of thermodynamic universality across different physical settings. While the critical exponent is universal, the amplitude $A_c$ is model-dependent, as shown in  Fig.~\ref{fig:R_tau_loglog}. To extract potentially universal geometric information, we define\footnote{
Similarly, one may define
\(
Q^{i}_{\;jk} \equiv A_c^{i}/(A_c^{j}A_c^{k}),
\)
and its permutations. However, these quantities do not yield any additional information beyond that encoded in the pairwise ratios \(Q^{i}_{\;j}\).
} the ratio of critical amplitudes
\begin{equation}
Q^i_{\;j} \equiv \frac{A_c^{i}}{A_c^{j}},
\qquad
i,j \in \{I,II,III\};
\end{equation}
where \(I\), \(II\), and \(III\) correspond to the metrics \(g^{I}\), \(g^{II}\), and \(g^{III}\), respectively. The behavior of \(Q^{i}_{\;j}\) for all models is displayed in Fig.~\ref{fig:Qtodos}. Although each critical amplitude exhibits a complicated logarithmic dependence on \(N\), this dependence cancels exactly in the ratios \(Q^{i}_{\;j}\) for all permutations of \(i\) and \(j\). The resulting expressions, while too cumbersome to be presented explicitly, reduce to polynomial functions of \(a/b\), thereby defining an intrinsic geometric property of the thermodynamic equilibrium space, as summarized in Table~\ref{tab:Ac_models}. This feature enables a direct connection with the Boyle temperature \(T_B\). As illustrated in Fig.~\ref{fig:Q}, all models display the same qualitative behavior. Since all ratios \(Q^{i}_{\;j}\) exhibit qualitatively similar behavior, we restrict our discussion, without loss of generality, to the simplest case, namely \(Q^{I}_{\;II}\).

\begin{table}[ht!]
\centering
\renewcommand{\arraystretch}{2.0}
\begin{tabular}{l|c|c|c}
\hline
\textbf{Model} & \boldmath{$T_*/T_B$} & \boldmath{$T_B/T_c$} & \boldmath{$T_*/T_c$} \\ \hline

\textbf{van der Waals} & $\approx 1.4$ & $\dfrac{27}{8}$ & $ 1.4 \times \dfrac{27}{8} \approx 4.73$ \\ \hline

\textbf{Berthelot} & $\approx 1.6$ & $\sqrt{\dfrac{27}{8}}$ & $1.6 \times \sqrt{\dfrac{27}{8}} \approx 2.94$ \\ \hline

\textbf{Redlich--Kwong} & $ \approx 1.4$ & $\approx 2.898$ & $ 1.4 \times 2.898 \approx 4.06$ \\ \hline

\end{tabular}
\caption{Characteristic ratios $T_*/T_B$ and $T_*/T_c$ for different fluid models. The condition $Q^{I}_{\;II}\to \infty$ occurs at $T_* \sim 1.4$--$1.6\,T_B$, corresponding to $T_* \sim 3$--$5\,T_c$ and depending on the model.}
\label{tab:ratio_Tstar_full}
\end{table}
In particular, a characteristic geometric temperature scale $T_* \sim (1.4\text{--}1.6)\,T_B$ emerges, at which the amplitude ratio $Q^{I}_{\;II} \to \infty$. As shown in Table~\ref{tab:ratio_Tstar_full} and Fig.~\ref{fig:Q}, this divergence consistently occurs at $T_*$ across all models, with $T_*$ lying close to the corresponding Boyle temperature. When expressed in units of the critical temperature, $T_* \sim (3\text{--}5)\,T_c$, indicating a nearly universal separation between the geometric scale and the critical point. Furthermore, the mathematical behavior of $Q^{I}_{\;II}$ is summarized in Table~\ref{tab:Q_structure_general}.

\begin{table}[h!]
\centering
\renewcommand{\arraystretch}{1.6}
\begin{tabular}{l|c}
\hline
\textbf{Regime of $Q^{I}_{\;II}$} 
& \textbf{Condition} \\ 
\hline

$Q^{I}_{\;II} \to \infty$ 
& $|A_c^{I}| \to \infty$  \\ 
\hline

$Q^{I}_{\;II} = 0$ 
& $A_c^{I} = 0$  \\ 
\hline

$Q^{I}_{\;II} = -1$ (local minimum) 
& $|A_c^{I}|,\, |A_c^{II}| \to \infty$  \\ 
\hline

$Q^{I}_{\;II} \to Q_{\infty}$ 
& $a/b \gg 1 \;\Rightarrow\; |A_c^{I}|,\, |A_c^{II}| \to 0$ \\

\multicolumn{2}{c}{
\small $Q_{\infty} = 3/5$ (vdW), $2/3$ (Bht), $\approx 0.65$ (R--K)
} \\ 
\hline

\end{tabular}
\caption{
Mathematical behavior of $Q^{I}_{\;II}$. 
The table summarizes its asymptotic regimes, showing that different fluid models 
share common limiting behaviors despite having distinct EoS.
}
\label{tab:Q_structure_general}
\end{table}
As shown in Table~\ref{tab:Q_structure_general} and Fig.~\ref{fig:Q}, in the asymptotic limit $a/b \gg 1$, the ratio $Q^{I}_{\;II}$ approaches a constant value, signaling the emergence of a universal regime in the geometric description. In this limit, both critical amplitudes $A_c^{I}$ and $A_c^{II}$ vanish; however, $A_c^{I}$ decays faster than $A_c^{II}$. As a result, their ratio stabilizes to a finite constant, revealing a hierarchical suppression between the amplitudes. Using the near-critical expansion of the scalar curvature given in Eq.~(\ref{Critical amp}), this behavior suggests that for $a/b \gg 1$ the leading divergent term in both scalars is suppressed, potentially giving way to a regular contribution of the form $\mathcal{R} \sim \mathcal{R}_0 + \mathcal{O}(\tau)$.

\begin{figure}[ht!]
    \centering
    \includegraphics[width=0.5\textwidth]{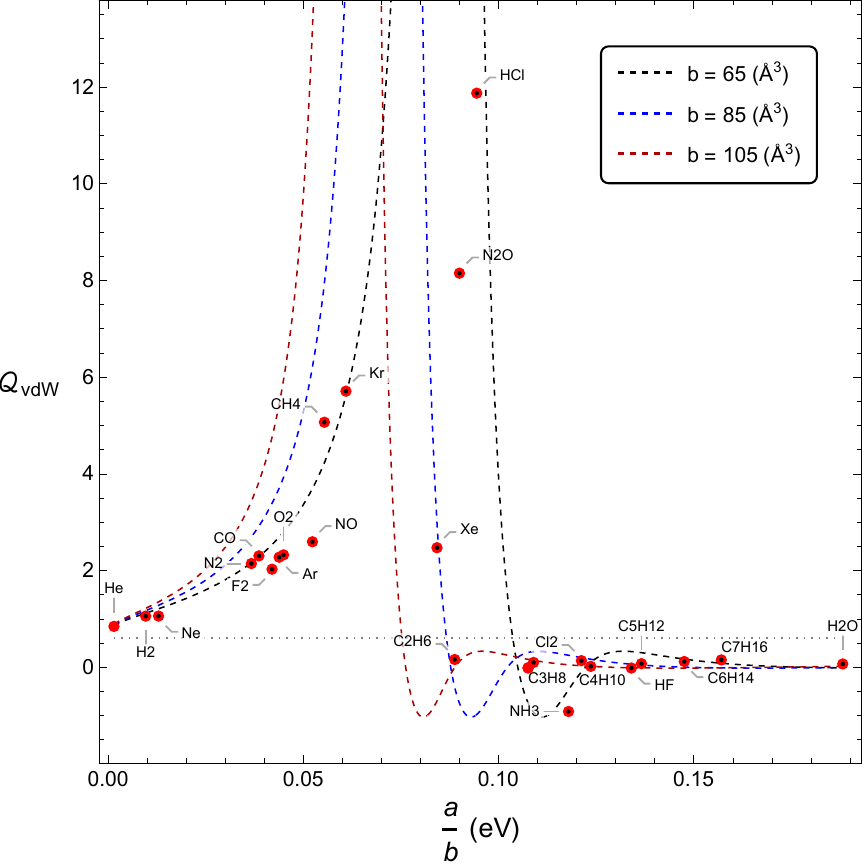}
\caption{
Critical ratio $Q^{I}_{\;II}$ for different molecular species within the vdW model. The dashed curves show $Q^{I}_{\;II}$ for different values of $b$, while the horizontal dotted black line indicates the asymptotic value $Q_{\infty}=3/5$. The colored points correspond to various molecular species, computed using the experimental values of $a$ and $b$ reported in \cite{johnston2014vdw}.
}

 \label{fig:Qvdw_real}
\end{figure}
Fig.~\ref{fig:Qvdw_real} shows that $Q^{I}_{\;II}$, although not a direct thermodynamic observable, behaves as a geometric quantity encoding the interplay between intermolecular interactions and excluded volume. It exhibits a non-monotonic dependence on $a/b$, with real fluids occupying a restricted region of the parameter space. Three regimes emerge: weakly interacting systems (e.g., H$_2$,He) at low $Q^{I}_{\;II}$; an intermediate regime where $Q^{I}_{\;II}$ attains a maximum (e.g., HCl, N$_2$O), corresponding to an optimal balance between attraction and excluded volume; and a suppressed regime with $Q^{I}_{\;II} \approx 0$, including both large molecules dominated by excluded volume (alkanes) and systems with directional interactions (e.g., H$_2$O).
\begin{table}[ht!]
\centering
\renewcommand{\arraystretch}{1.6}
\begin{tabular}{l|c|c|c|c}
\hline
\textbf{Model} & \boldmath{$Q_{\max}$} & \boldmath{$T_{\max}/T_B$} & \boldmath{$Q_{\min}$} & \boldmath{$T_{\min}/T_B$} \\ \hline

\textbf{van der Waals} 
& $\approx 0.3$ 
& $\approx 2.2$ 
& $-1$ 
& $\approx 1.8$ \\ \hline

\textbf{Berthelot} 
& $\approx 2.27$ 
& $\approx 0.64$ 
& $-1$ 
& $\approx 1.0$ \\ \hline

\textbf{Redlich--Kwong} 
& $\approx 4.2$ 
& $\approx 1.0$ 
& $-1$ 
& $\approx 1.22$ \\ \hline

\end{tabular}
\caption{Location of the extrema of  $Q^{I}_{\;II}$ for different models. While both the position and value of the maximum depend on the model, the minimum consistently occurs at $Q^{I}_{\;II} = -1$. The parameter $a/b$ is expressed in units of the respective Boyle temperature.}
\label{tab:Q_extrema}
\end{table}
The extrema of $Q^{I}_{\;II}$ are summarized in Table~\ref{tab:Q_extrema}, where again a geometric temperature scale is used to facilitate comparison with the corresponding Boyle temperature $T_B$ of each model. The extrema occur at characteristic values of $T/T_B$, highlighting that $T_B$ acts as a natural geometric scale of the system. Remarkably, the local minimum of $Q^{I}_{\;II}$ attains the universal value $Q_{\min} = -1$ across all considered models. This implies that at $T_{\min}$
\begin{equation}
\left.Q^{I}_{\;II} \right|_{T_{\min}}=\left.\frac{A_c^{I}}{A_c^{II}}\right|_{T_{\min}} = -1 
\end{equation}
This result indicates an emergent symmetry in the GTD description, independent of the specific model. It would be particularly interesting to examine whether this purely geometric factor $Q^{i}_{\;j}$ remains independent of system size in gravitational settings, such as black holes, where extensivity is subtle. 
Establishing such independence could further support the interpretation of these amplitude ratios as universal geometric features beyond standard thermodynamic systems.

\section{Zero-Curvature Curves in Geometrothermodynamics}
\label{Zero curv}
\begin{figure}[ht!]
\centering

\begin{minipage}{0.325\textwidth}
    \centering
    \includegraphics[width=\linewidth]
    {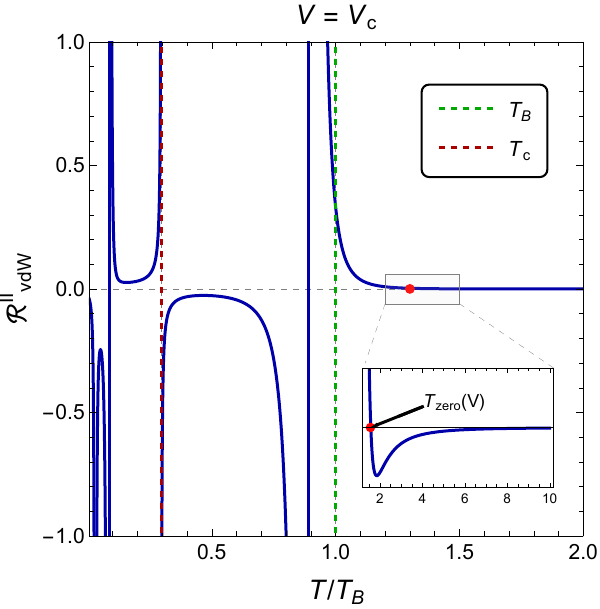}
    
    (a)
\end{minipage}
\hfill
\begin{minipage}{0.315\textwidth}
    \centering
    \includegraphics[width=\linewidth]
    {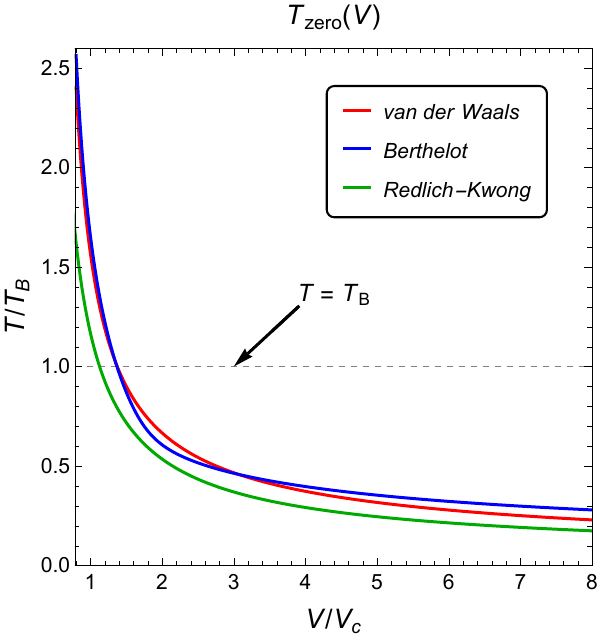}
    
    (b)
\end{minipage}
\hfill
\begin{minipage}{0.325\textwidth}
    \centering
    \includegraphics[width=\linewidth]
    {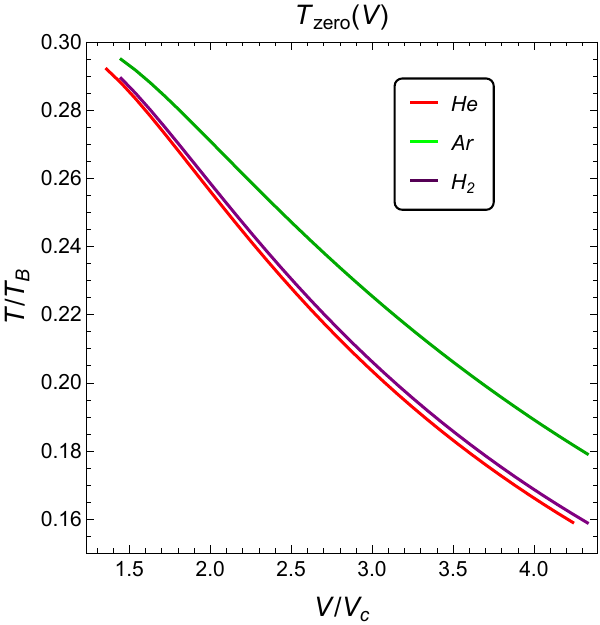}
    
    (c)
\end{minipage}
    
\caption{
(a) $\mathcal{R}^{\mathrm{II}}$ for the vdW fluid obtained from the free-energy thermodynamic potential, Eq.~\eqref{Free vdW}, with $a=b=N=k_B=1$. 
Panels (b) and (c) display the corresponding $T_{\mathrm{zero}}(V)$ profiles. 
Panel (b) compares the vdW, Berthelot, and Redlich--Kwong fluid models in normalized units with $a=b=N=k_B=1$, while panel (c) presents the vdW model using experimental parameters $(a,b)$ for Helium, Argon, and Hydrogen.
}
    \label{fig:TzeroBranches}
\end{figure}

In this section, we investigate the zero-curvature structure associated with the GTD scalar curvature for different real fluid EoS. For simplicity, we restrict the analysis to the scalar curvature $\mathcal{R}^{II}$ computed from the Helmholtz free-energy representation, following the same setup introduced in Section~\ref{universa}, while the analysis of $\mathcal{R}^{I}$ and $\mathcal{R}^{III}$  will be left for future work. Since the scalar curvature $\mathcal{R}^{II}$ may exhibit multiple distinct zeros, here we focus only on the zero located near the Boyle temperature scale, which appears to encode the relevant universal geometric behavior. The corresponding condition is defined by
\begin{equation}
    \mathcal{R}^{II}(V,T,N;a,b)=0.
\end{equation}
For each fluid model, we track the zero of the GTD scalar curvature as the thermodynamic volume varies, thereby defining the function $T_{\mathrm{zero}}(V)$ shown in Fig.~\ref{fig:TzeroBranches}(a). Fig.~\ref{fig:TzeroBranches}(b) displays the numerical $T_{\mathrm{zero}}(V)$ profiles for the vdW, Berthelot, and Redlich--Kwong models in normalized units, whereas Fig.~\ref{fig:TzeroBranches}(c) shows the corresponding behavior for using experimental parameters taken from Ref.~\cite{johnston2014vdw}.

\begin{figure}[ht!]
    \centering
    
   \begin{minipage}{0.32\textwidth}
    \centering
    \includegraphics[width=\linewidth]
    {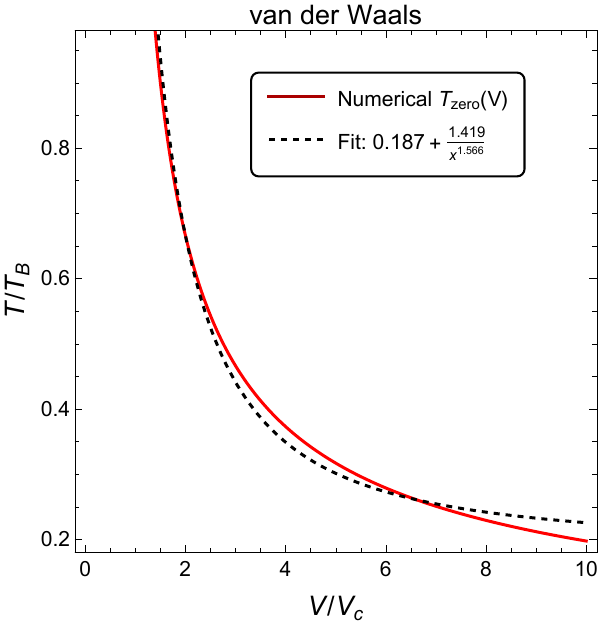}
    
    (a)
\end{minipage}
\hfill
\begin{minipage}{0.325\textwidth}
    \centering
    \includegraphics[width=\linewidth]
    {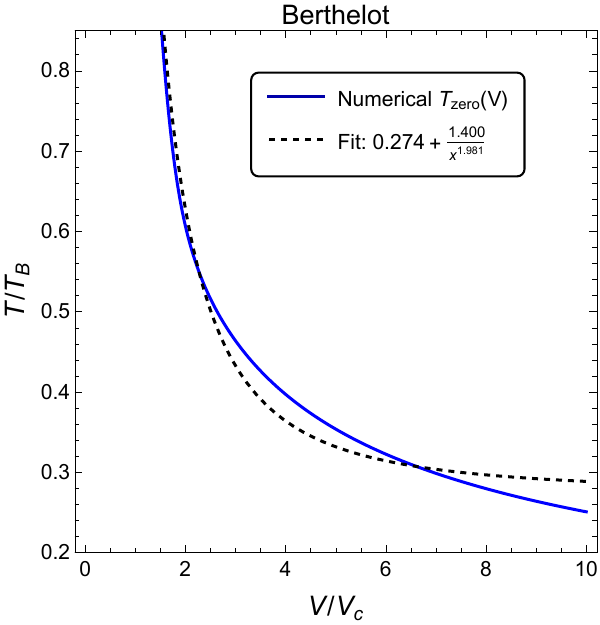}
    
    (b)
\end{minipage}
\hfill
\begin{minipage}{0.325\textwidth}
    \centering
    \includegraphics[width=\linewidth]
    {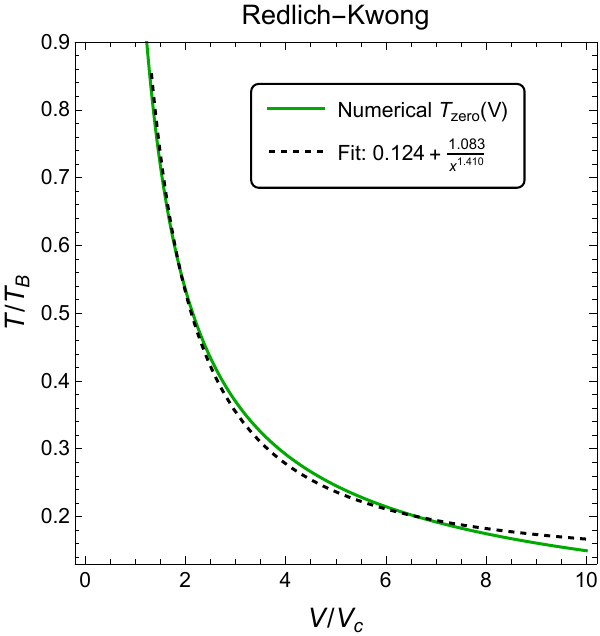}
    
    (c)
\end{minipage}
    
    \caption{
    Power-law fits of the 
    $T_{\mathrm{zero}}(V)$ function. 
    (a) van der Waals model. 
    (b) Berthelot model. 
    (c) Redlich--Kwong model. 
    The solid curves correspond to the numerical GTD data, while 
    the dashed black curves represent the best-fit functions of the form
    $
    T_{\mathrm{zero}}/T_B
    =
    A_0(V/V_c)^{-\gamma}+C
    $.
    }
    
    \label{fig:TzeroPowerFits}
\end{figure}
Interestingly, the different fluid models exhibit a smooth monotonic decreasing behavior as the thermodynamic volume increases. To characterize this behavior quantitatively, we fitted the numerical GTD data using the scaling relation
$
T_{\mathrm{zero}}/T_B
=
A_0(V/V_c)^{-\gamma}+C,
$
where $A_0$ determines the amplitude of the power-law contribution, $\gamma$ is the corresponding scaling exponent, and $C$ represents the asymptotic large-volume value of $T_{\mathrm{zero}}/T_B$. The corresponding fits are displayed in Fig.~\ref{fig:TzeroPowerFits}, where the dashed curves represent the best-fit power-law profiles for the vdW, Berthelot, and Redlich--Kwong models. As shown in Fig.~\ref{fig:TzeroPowerFits}, the fitted exponents $\gamma$ differ for each fluid model, indicating that the asymptotic GTD scaling retains information about the specific thermodynamic structure encoded in the corresponding EoS. In particular, the Berthelot fluid exhibits an exponent close to $\gamma \simeq 2$, while the vdW and Redlich--Kwong fluids display $\gamma \simeq 3/2$. The parameter $C$ also varies among the models, suggesting that the large-volume limit is not universal. In the following subsection, we use a Bayesian inference analysis of these parameters in order to further test the statistical robustness and stability of the GTD scaling analysis.

% fitting parameters are summarized in Table~\ref{tab:powerlawfits}. The quality of the fits is quantified through the coefficient of determination
%\begin{equation}
 %   R^2 = 1-\frac{\sum_i \left(y_i-\hat{y}_i\right)^2}
%{\sum_i \left(y_i-\bar{y}\right)^2},
%\end{equation}
%where \(y_i\) denote the numerical data, \(\hat{y}_i\) the fitted values, and \(\bar{y}\) the mean of the dataset. The obtained values, \(R^2 \gtrsim 0.998\), indicate an excellent agreement between the numerical results and the proposed scaling law, providing strong evidence for an emergent power-law behavior. It would be interesting to explore whether the remaining zeros of \(\mathcal{R}^{II}\) obey similar scaling relations or instead display qualitatively different asymptotic behaviors.

\begin{comment}

\begin{table}[H]
\centering

\begin{tabular}{c c c c c}
\hline
Model & $A_0$ & $\gamma$ & $C$ & $R^2$ \\
\hline

vdW 
& $1.41925 \pm 0.00505$ 
& $1.56582 \pm 0.01207$ 
& $0.18717 \pm 0.00265$ 
& $0.997806$ \\

Berthelot 
& $1.40022 \pm 0.00487$ 
& $1.98081 \pm 0.01341$ 
& $0.27393 \pm 0.00192$ 
& $0.998068$ \\

Redlich--Kwong 
& $1.08271 \pm 0.00299$ 
& $1.40991 \pm 0.00775$ 
& $0.12445 \pm 0.00176$ 
& $0.998809$ \\

\hline
\end{tabular}
\caption{
Power-law fitting parameters of
$T_{\mathrm{zero}}(V)$ obtained from the condition $\mathcal{R}^{II}=0$. 
}
\label{tab:powerlawfits}
\end{table}
\end{comment}

%===========================================================

%============================================================
\subsection{Bayesian Analysis of the GTD Zero-Curvature Curve}
%============================================================
Next, we performe a Bayesian parameter inference using MCMC methods on the numerical reconstruction of the zero-curvature curve $T(V)$ obtained from the condition $\mathcal{R}^{II}=0$. The resulting scaling relation was modeled as
\begin{equation}
\frac{T_{\mathrm{zero}}}{T_B}
=
A_0
\left(
\frac{V}{V_c}
\right)^{-\gamma}
+
C. \label{model}
\end{equation}
Within the Bayesian framework, the posterior probability distribution of the parameter vector
$\vartheta=(A_0,\gamma,C)$ is obtained from Bayes' theorem,
\begin{equation}
P(\vartheta|D)
\propto
\mathcal{L}(D|\vartheta)\,\Pi(\vartheta),
\end{equation}
where $D$ represents the numerical GTD data,
\(
\mathcal{L}(D|\vartheta)
\)
is the likelihood function, and
\(
\Pi(\vartheta)
\)
denotes the prior distribution for the parameter vector
\(
\vartheta
\). Assuming independent Gaussian uncertainties for the numerical data, each data point was modeled as

\begin{equation}
\left(\frac{T}{T_B}\right)_i^{\mathrm{(num)}}
\sim
\mathcal{N}
\left(
\left(\frac{T}{T_B}\right)_i^{\mathrm{(model)}},
\,\sigma_i^2
\right),
\end{equation}
where
\(
\mathcal{N}(\mu,\sigma^2)
\)
denotes a Gaussian distribution with mean
\(
\mu
\)
and variance
\(
\sigma^2
\).
Consequently, the likelihood function was defined as

\begin{equation}
\mathcal{L}(D|\vartheta)
\propto
\exp\left[
-\frac{1}{2}
\sum_i
\left(
\frac{
\left(\frac{T}{T_B}\right)_i^{\mathrm{(num)}}
-
\left(\frac{T}{T_B}\right)_i^{\mathrm{(model)}}
}{
\sigma_i
}
\right)^2
\right].
\end{equation}
where
\(
T_i^{\mathrm{(num)}}
\)
corresponds to the numerical GTD zero-curvature temperatures expressed in reduced variables, while the theoretical model is given by Eq.~\eqref{model}, and $\sigma_i$ represents the numerical uncertainty associated with each data point. Non-informative uniform priors were adopted for all parameters within physically reasonable intervals
\begin{equation}
\Pi(A_0,\gamma,C)
=
\mathcal{U}(A_0)\,
\mathcal{U}(\gamma)\,
\mathcal{U}(C),
\end{equation}
\begin{figure}[ht!]
    \centering

    \begin{minipage}{0.495\textwidth}
        \centering
        \includegraphics[width=\textwidth]{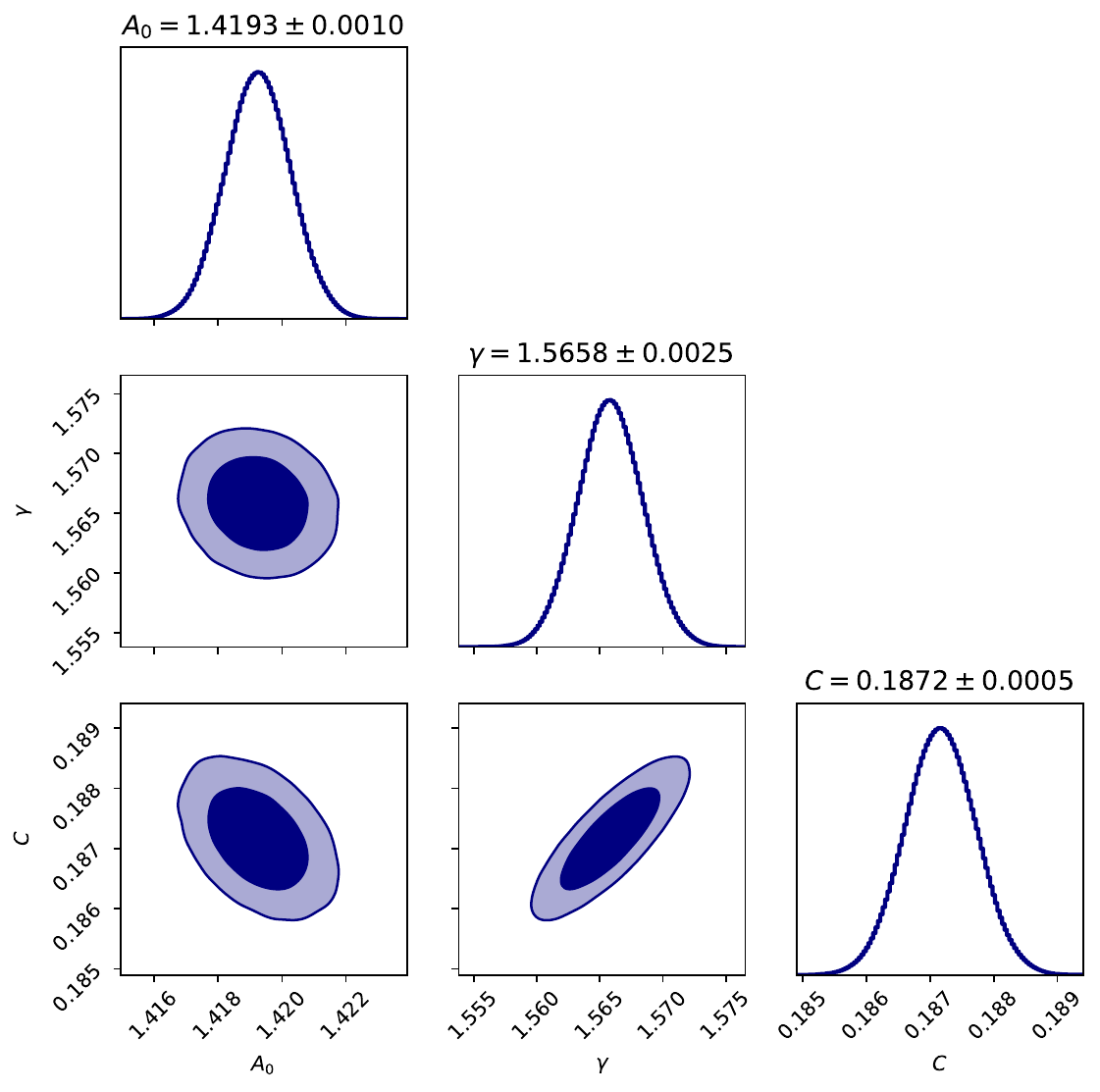}
        
        (a) 
    \end{minipage}
    \hfill
    \begin{minipage}{0.495\textwidth}
        \centering
        \includegraphics[width=\textwidth]{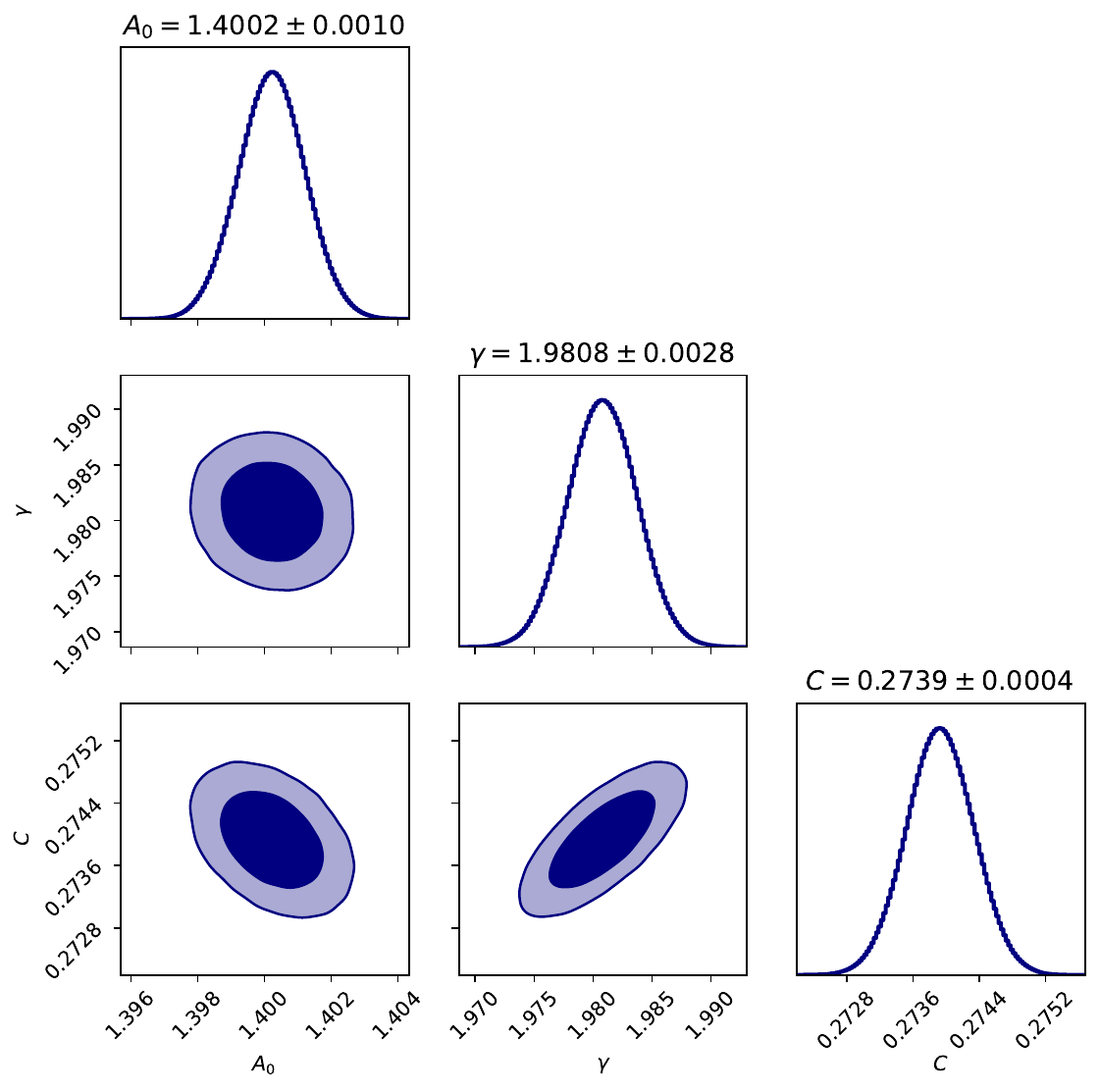}
        
        (b) 
    \end{minipage}

    \vspace{0.3cm}

    \begin{minipage}{0.495\textwidth}
        \centering
        \includegraphics[width=\textwidth]{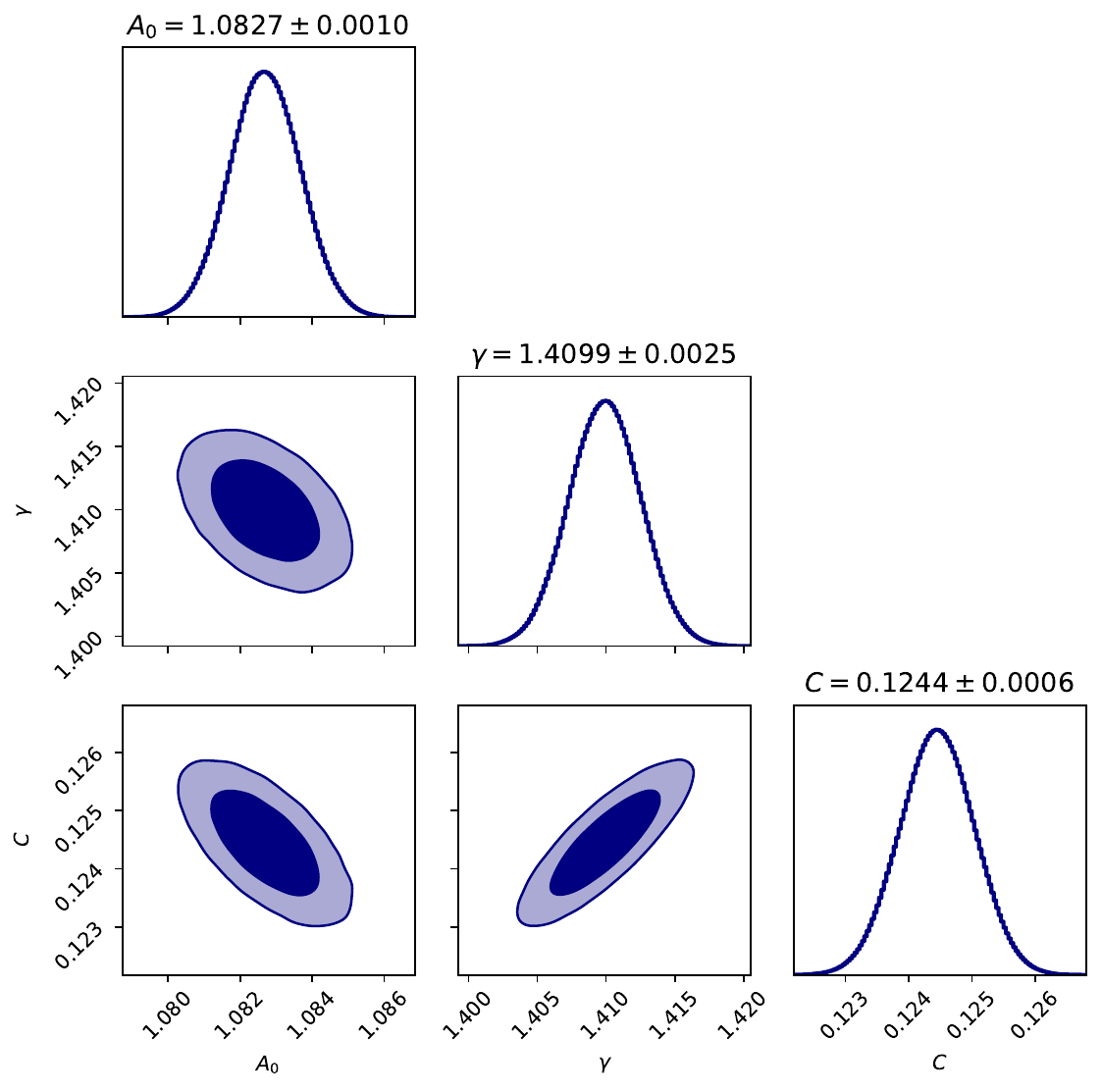}
        
        (c) 
    \end{minipage}

    \caption{Posterior corner plots for the Bayesian inference of the GTD scaling law parameters $(A_0,\gamma,C)$ for the (a) vdW, (b) Berthelot, and (c) Redlich--Kwong models.}
    \label{fig:cornerplots}
\end{figure}
where $\mathcal{U}$ denotes a uniform probability distribution, corresponding to flat priors assigned within physically admissible parameter ranges, thereby avoiding any strong prior bias in the reconstruction of the scaling law. The posterior distributions were sampled using the affine-invariant ensemble sampler implemented in the Python package \texttt{emcee}. Multiple walkers were evolved over sufficiently long chains to ensure proper convergence and efficient exploration of the parameter space. The resulting posterior distributions are sharply peaked and approximately Gaussian, indicating stable Bayesian reconstructions and well-constrained parameter estimation. Fig.~\ref{fig:cornerplots} displays the posterior corner plots obtained for the vdW, Berthelot, and Redlich--Kwong models. The diagonal panels show the marginalized posterior distributions of each parameter, while the off-diagonal panels display the corresponding two-dimensional confidence regions. The ellipses represent the $1\sigma$, $2\sigma$, and $3\sigma$ credible regions, illustrating the statistical correlations between the inferred parameters. Overall, the posterior distributions remain well localized with no evidence of multimodality, indicating stable Bayesian reconstructions for all thermodynamic models. 
\begin{figure}[ht!] 
\centering \includegraphics[width=0.6\linewidth]{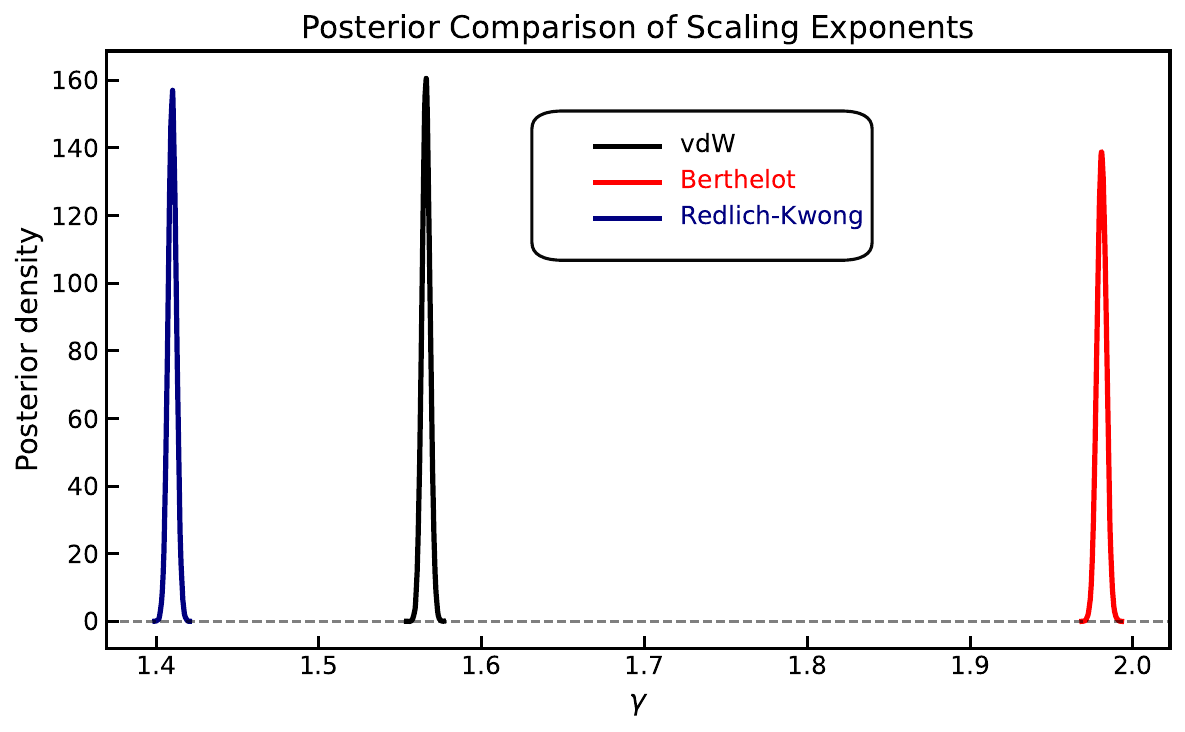} \caption{Posterior distributions of the scaling exponent $\gamma$ for the vdW, Berthelot, and Redlich--Kwong models.} \label{fig:gammaComparison} \end{figure}
Additionally, Fig.~\ref{fig:gammaComparison} compares the posterior distributions of the scaling exponent $\gamma$. The narrow posterior profiles further support the robustness of the scaling behavior. Finally, to quantify the statistical dependence among the inferred parameters, we computed the posterior correlation matrix. For the vdW model, the resulting matrix is given by

\begin{equation}
\mathrm{Corr}=
\begin{pmatrix}
1 & -0.139 & -0.418 \\
-0.139 & 1 & 0.851 \\
-0.418 & 0.851 & 1
\end{pmatrix}.
\end{equation}
The correlation analysis reveals a strong positive correlation between the scaling exponent $\gamma$ and the constant offset $C$, indicating a partial statistical degeneracy between the effective decay rate and the asymptotic shift of the zero-curvature temperature curve. In contrast, the amplitude parameter $A_0$ remains only weakly correlated with the remaining parameters, suggesting that the overall normalization of the reconstructed \(T_{\mathrm{zero}}\) relation is comparatively stable. The median posterior values and corresponding $1\sigma$ credible intervals are summarized in Table~\ref{tab:bayesianfit}.

\begin{table}[ht!]
\centering

\begin{tabular}{c c c c}
\hline\hline

Model & $A_0$ & $\gamma$ & $C$ \\

\hline

van der Waals
&
$1.4193 \pm 0.0010$
&
$1.5658 \pm 0.0025$
&
$0.1872 \pm 0.0005$

\\

Berthelot
&
$1.4002 \pm 0.0010$
&
$1.9808 \pm 0.0028$
&
$0.2739 \pm 0.0004$

\\

Redlich--Kwong
&
$1.0827 \pm 0.0010$
&
$1.4099 \pm 0.0025$
&
$0.1244 \pm 0.0006$

\\

\hline\hline
\end{tabular}
\caption{
Bayesian posterior estimates for the parameters of the reduced zero-curvature curve \(T_{\mathrm{zero}}/T_B\).
}\label{tab:bayesianfit}
\end{table}
Overall, the Bayesian analysis confirms the stability of the zero-curvature scaling relation across the different thermodynamic fluids. The narrow posterior distributions, stable MCMC chains, and consistent credible intervals collectively indicate that the inferred \(T_{\mathrm{zero}}\) curve is statistically robust and insensitive to sampling fluctuations. Moreover, the posterior distributions obtained for the different thermodynamic models show essentially no overlap, particularly for the scaling exponent $\gamma$ Fig.~\ref{fig:gammaComparison}. This strongly suggests that the differences among the inferred parameters are not numerical artifacts of the reconstruction procedure, but rather reflect genuine distinctions in the thermodynamic structure encoded by each EoS. Consequently, the reconstructed zero-curvature curve retains nontrivial information about the underlying fluid model.

%============================================================

%To verify the convergence and stability of the MCMC chains, we computed the trace plots shown in Fig.~\ref{fig:traceplots}.

%\begin{figure*}[t!]
 %  \centering
  % \includegraphics[width=0.32\textwidth]{vdW_chains.pdf}
   % \includegraphics[width=0.32\textwidth]{Berthelot_chains.pdf}
    %\includegraphics[width=0.32\textwidth]{RedlichKwong_chains.pdf}
    %\caption{Trace plots for the posterior sampling chains corresponding to the GTD scaling parameters.}
    %\label{fig:traceplots}
%\end{figure*}

%The chains exhibit stable stationary behavior and efficient exploration of the posterior parameter space, indicating that the MCMC sampling reached convergence without visible long-term drift or multimodal transitions.

\section{Conclusions}
\label{sec:VI}
In this work, we developed a unified entropic framework that incorporates a broad class of real fluid models within a common thermodynamic description. The formalism is governed by two functions, $A(V,N,T)$ and $\Theta(T)$, which encode the effects of intermolecular interactions and thermal corrections while preserving thermodynamic consistency through the associated integrability condition. Different choices of these functions reproduce well-known EoS, including the van der Waals, Berthelot, Redlich--Kwong, and Peng--Robinson models, demonstrating that a variety of phenomenological fluid descriptions can be embedded into a common thermodynamic structure.

The entropy representation provides a natural setting for the geometric description of thermodynamics. Since all fluid models considered in this work are homogeneous, they satisfy the Euler identity and admit a direct GTD description. Within this framework, curvature singularities reproduce the critical behavior associated with phase transitions. First, in Section~\ref{GTD}, we analyzed the equilibrium space of the vdW model using the GTD scalar curvature $\mathcal{R}^{II}$. Under the interaction hypothesis, the sign of $\mathcal{R}^{II}$ encodes the nature of the effective thermodynamic interactions, whereas its magnitude $|\mathcal{R}^{II}|$ measures their intensity. Accordingly, the gaseous phase exhibits weaker interactions than the liquid phase, in agreement with previous results obtained for black holes and other gravitational thermodynamic systems \cite{ladino2025phase,Ladino:2024ned,romero2026quasi}. This behavior is particularly evident for the subcritical isobar $P<P_c$, where the magnitude of the curvature decreases as the temperature increases, indicating progressively weaker interactions as the gaseous phase approaches the ideal-gas regime. Furthermore, in Section~\ref{universa}, we extended the GTD analysis to all fluid models by employing the Helmholtz free energy and the three GTD metrics. Remarkably, all of them reproduce the phase structure of the models considered. In addition, we found that the GTD curvature is characterized by a universal critical exponent $\zeta=1$, independently of both the fluid model and the GTD metric, whereas the critical amplitudes $A_c$ remain model- and metric-dependent. This observation naturally led us to introduce the amplitude ratio $Q^{i}_{\;j}$ as a novel dimensionless quantity for characterizing critical phenomena through the relative strength of curvature divergences. Unlike the critical amplitudes themselves, $Q^{i}_{\;j}$ is independent of the system size $N$ and therefore provides an intrinsic measure of criticality. Remarkably, all fluid models considered exhibit values within a narrow range and, when expressed in terms of the Boyle temperature, the corresponding divergences occur in the interval $T \simeq (1.4-1.6)\,T_B$. Using experimental values of the parameters $a$ and $b$ \cite{johnston2014vdw}, we further showed that different molecular species can be organized according to their corresponding $Q^{i}_{\;j}$ values. Together, these results suggest that $Q^{i}_{\;j}$ captures universal aspects of critical behavior and may provide the basis for a classification scheme applicable not only to real fluids but also to more general thermodynamic systems, including black holes and other quasi-homogeneous systems.\\

Finally, in Section~\ref{Zero curv}, the geometric significance of the zero-curvature condition was further explored by reconstructing the locus of temperatures for which $\mathcal{R}^{II}=0$. Within the interaction hypothesis adopted throughout this work, a vanishing GTD curvature identifies a thermodynamic state of effective interaction neutrality, where the attractive and repulsive contributions exactly balance each other. From this perspective, the condition $\mathcal{R}=0$ plays a role analogous to that of the ideal-gas limit, but within interacting systems, defining a non-trivial geometric boundary between regions characterized by different interaction regimes. The existence of well-defined and model-dependent curves suggests that they may encode physically relevant thermodynamic information rather than being mere mathematical artifacts of the GTD formalism. The extent to which these structures encode signatures of the underlying statistical description, as well as how similar features may emerge within the GTD metrics $g^{I}$ and $g^{III}$, remains an open question. To gain further insight into these geometric signatures, we proposed a polynomial ansatz for the zero-curvature temperature curve associated with $\mathcal{R}^{II}$ and reconstructed it through a Bayesian MCMC analysis. The posterior distributions exhibit narrow confidence regions, confirming the stability of the proposed scaling relations across the different fluid models considered. Moreover, the reconstructed parameters remain clearly separated from one model to another, indicating that the corresponding zero-curvature curves retain distinct thermodynamic signatures. Clarifying the origin of these differences may shed light on the statistical origin of the fluid models under consideration. It would also be worthwhile to apply the present framework and Bayesian reconstruction scheme to ideal quantum gases, whose GTD description exhibits non-trivial curvature effects associated with quantum statistics and Bose--Einstein condensation \cite{zaldivar2023ideal}. We leave these investigations for future work.

\section*{Acknowledgments}
CRF and JML acknowledge support from Conahcyt-Mexico, grants No. 4003366 and No. 402076. This work was supported by UNAM-DGAPA-PAPIIT, grant No. 108225 and Conahcyt, grant No. CBF-2025-I-243. SZ acknowledge support from SECIHTI grant Estancias Posdoctorales por M\'exico.
\renewcommand{\addcontentsline}[3]{\oldaddcontentsline{#1}{#2}{#3}}
\appendix

\let\oldaddcontentsline\addcontentsline
\renewcommand{\addcontentsline}[3]{}
 \setlength{\bibsep}{0pt}
\bibliographystyle{unsrt}
\bibliography{referenciasGTD}
\end{document}